\documentclass[%
reprint,
superscriptaddress,
 amsmath,amssymb,
 aps,
prb,
floatfix,
]{revtex4-2}
\usepackage{libertine}
\usepackage{graphicx}
\usepackage{dcolumn}
\usepackage{bm}
\usepackage{hyperref}

\usepackage{algorithmic}
\usepackage{changes}

\begin{document}


\title{\sc \Large Illustrated tutorial on global optimization in nanophotonics}
\author{Pauline Bennet}
\author{Denis Langevin}
\author{Chaymae Essoual}
\affiliation{Universit\'e Clermont Auvergne, Clermont Auvergne INP, CNRS, Institut Pascal, F-63000 Clermont-Ferrand, France}
\author{Abdourahman Khaireh-Walieh}
\affiliation{LAAS, Universit\'e de Toulouse, CNRS, Toulouse, France}
\author{Olivier Teytaud}
\affiliation{Meta AI Research Paris, France}
\author{Peter Wiecha}
\affiliation{LAAS, Universit\'e de Toulouse, CNRS, Toulouse, France}
\author{Antoine Moreau}
\affiliation{Universit\'e Clermont Auvergne, Clermont Auvergne INP, CNRS, Institut Pascal, F-63000 Clermont-Ferrand, France}
\email{antoine.moreau@uca.fr}

\date{\today}

\begin{abstract}
Numerical optimization for the inverse design of photonic structures is a tool which is providing increasingly convincing results -- even though the wave nature of problems in photonics makes them particularly complex. In the meantime, the field of global optimization is rapidly evolving but is prone to reproducibility problems, making it harder to identify the right algorithms to use. This paper is thought as a tutorial on global optimization for photonic problems. We provide a general background on global optimization algorithms and a rigorous methodology for a physicist interested in using these tools -- especially in the context of inverse design. We suggest algorithms and provide explanations for their efficiency. We provide codes and examples as an illustration than can be run online, integrating quick simulation code and Nevergrad, a state-of-the-art benchmarking library. Finally, we show how physical intuition can be used to discuss optimization results and to determine whether the solutions are satisfactory or not.
\end{abstract}

\maketitle

\section{Introduction}

While efficient automated design methods for multilayered structures have emerged in the 1970s, typically, numerical optimization has been used only more recently, thanks to the increase in the available computational power and the progress in simulation techniques. 
These developments lead to methods providing original and efficient designs for three-dimensional structures, for which no design rules exist\cite{molesky2018inverse, campbell2019review, elsawy2020numerical,chen2022algorithm}. In photonics, the most promising approaches so far are inspired by successful methods from mechanics and are based on local optimization algorithms\cite{bendsoe2003topology}. However, in photonics, the wave nature of the problem typically generates a large number of local minima, making global optimization algorithms valuable tools, while they are in many cases unreasonably expensive in mechanics\cite{sigmund2011usefulness}. 

Numerical optimization is a domain in which significant progress has been made in the last two decades, with enormous practical implications. Recent results suggest that modern global optimization algorithms are able to provide us with original and efficient photonic designs\cite{barry2020evolutionary} that are particularly convincing as they can be understood from a physical point of view\cite{bennet2021analysis}. However, reproducibility problems have made the important results in the field optimization harder to identify and to trust\cite{sorensenmetaheuristicsmetaphorexposed2015} -- especially for researchers in other communities.

The aim of this paper is to serve as a tutorial for photonics specialists who are interested in using modern numerical optimization tools. We provide insights, practical tips, and guidance to help researchers navigate the challenges and pitfalls associated with optimization. We demonstrate how simulation tools and state-of-the-art optimization libraries can be easily integrated to effectively tackle inverse design problems. 

Specifically, we provide examples of multi-layer photonics problems, simulated with the PyMoosh toolkit \cite{pymoosh} and optimized using the Nevergrad python library\cite{nevergrad}.
We present a comprehensive methodology that includes defining relevant observables, choosing optimization strategies, and computing specific criteria to assess the reliability of the obtained solutions. We offer practical examples inspired by real-world problems involving multilayered structures, but we have also 
included the optimization of a 2D optical grating and a 3D plasmonic structures to show that these technique can be applied even in the most complex setups. These examples effectively illustrate our methodology and make it easy to transpose to other situations. For easy reproducibility, the codes are providedin an online repository\mbox{\cite{bennet_2023_10228377}}.

In the first part, we provide essential background information on automated design of photonic structures and optimization. We also introduce the test cases that we use to demonstrate the concepts discussed in this paper.

The second section provides an overview of algorithm categories, explains in detail the inner workings of some algorithms, and outlines key observations to make during algorithm execution, as these observations provide insights into the success or failure of the optimization. 

In the third part, we walk through the steps of running an optimization, showcase the benchmarking of algorithms using our examples, and provide criteria for selecting the most effective algorithms. Additionally, we conduct a thorough physical analysis of our solutions to evaluate their quality. 

In the final part, we present general guidelines for optimizing photonic structures using global optimization algorithms. These guidelines serve as a methodology to avoid common pitfalls and maximize the potential of the optimization process.

\section{General background and description of the test cases} 

In the fields of photonics and optimization, numerous concepts and terminologies have been introduced over the years. As we now require tools and concepts from both domains to optimize complex photonic structures, it is important to present an overview of the vocabulary that has been developed and provide clear definitions for key terms.

\subsection{Fundamentals of optimization in photonics}

\begin{figure*}
    \centering
    \includegraphics[width=\linewidth]{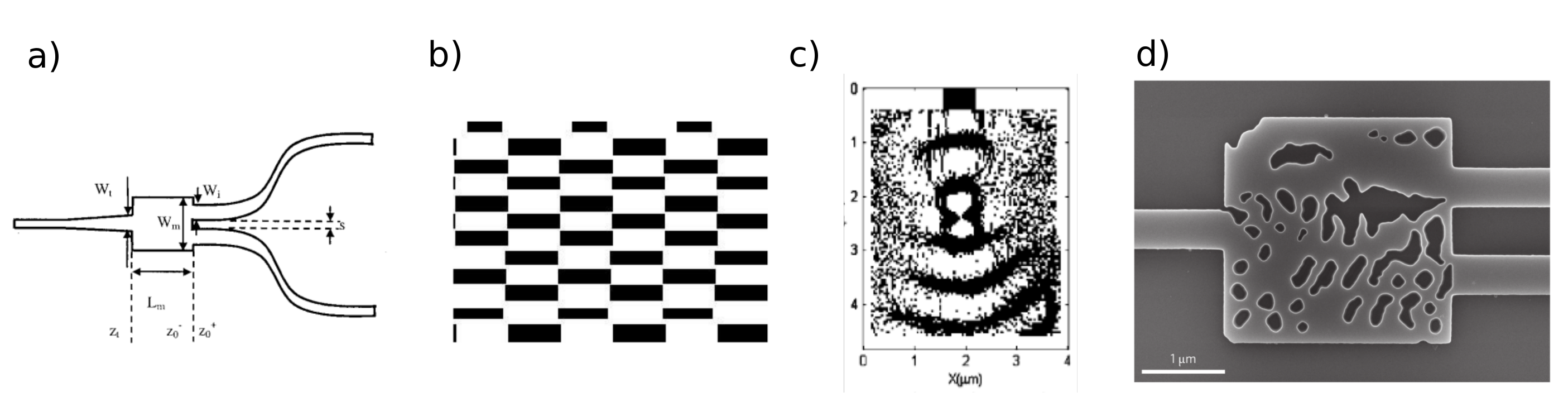}
    \caption{\textbf{Parametrization.} The process of selecting the parameters to describe the structure plays a crucial role in determining the types of solutions that can be obtained.
    (a) Less than 10 parameters are optimized. The geometry of the structure is fully constrained. We refer to problems of such low degrees of freedom as optimization, but not inverse design. Image from \cite{wang2002optimal}.
    (b) Around a hundred parameters are optimized. While the geometry is constrained since the structure is periodic, the blocks can be any size and position in the period. Optimization could have produced a Bragg mirror as well as a diffraction grating. Here, an intermediate structure is produced presenting characteristics from both. Given the wide range of possibilities, this can be deemed an inverse problem. Image from \cite{barry2020evolutionary}.
    (c) Around a thousand parameters are optimized on a pixel-based optimization. Each pixel is filled with one material or another to create the final design. Image from \cite{gondarenko2006spontaneous}
    (d) Tens of thousands of parameters are optimized. We  call this type of image parametrization topology optimization. Image from \cite{piggott2015inverse}.}
    \label{fig:param}
\end{figure*}

{\bf Defining the optimization problem.} To apply optimization techniques for improving a photonic structure, the structure must be represented by a finite set of parameters. This process of selecting the parameters to describe the structure is known as \textit{parametrization}, and it plays a crucial role in determining the types of solutions that can be obtained, potentially introducing biases.
Figure~\ref{fig:param} presents typical examples of problems with increasing complexity in their parametrization.

In photonics, many problems are continuous, meaning that valid structures can be achieved by making the parameters change continuously. In the present paper we focus on continuous optimization. However, when the problem is not continuous, it is said to be discrete or combinatorial, requiring specialized algorithms\cite{relengler, doerr2021survey, back2023evolutionary}. As will be explained below, discrete problems in photonics are often made continuous in order to leverage  gradient-based approaches. We underline that this comes at the cost of extra complications and that discrete algorithms can also provide interesting solutions~\cite{gagnon2013multiobjective, teytaud2022discrete}, even if they are less often considered in photonics\cite{campbell2019review}. 

It is then necessary to define a \textit{cost function} that quantifies the discrepancy between the actual optical response of the structure and the desired response. This cost function serves as a measure of how close the performance of the structure is to the target, even if this is not a measure in a strict mathematical sense. Finally, an optimization domain must be defined, typically by setting bounds on the parameter values. Together, the cost function and the optimization domain form the optimization problem.
The \textit{global optimum} is the point of the optimization domain where the value of the cost function is the lowest. While it is simple to determine whether a solution corresponds to a minimum of the cost function, called a local minimum, it is generally impossible to know whether such a minimum is the global optimum, {\em i.e.,} the solution with the lowest possible value of the cost function, and hence the closest to the desired response. The surface representing the cost function as a function of the parameters is called the cost function landscape.

We underline it is often useful to put limits on the values of the cost function. A lot of cost functions in photonics are based on reflection or transmission coefficients, so that the physical limits put on such coefficients translate into physical boundaries for the cost function. Values outside these boundaries may be indicative of a numerical error, something algorithms often tend to find and exploit.

An attraction basin is the region of the optimization domain for which a solution will be found whenever a local algorithm starts in this region. An optimization domain can typically be divided into attraction basins.

{\bf Global optimization and Black Box Optimization.} The search for a global optimum is called global optimization (GO). Algorithms that optimize without using the gradient or other side information, are called Black Box Optimization (BBO). BBO and GO are not synonymous: the former refers to the absence of additional information (gradient or other), whereas the latter refers to caring about global minima rather than local minima. There is a large overlap between the two categories of algorithms though, even if some BBO algorithms can be deemed local and if some GO algorithms exploit the information provided by the gradient\cite{wu2017bayesian,garcia2021bayesian}.

BBO algorithms are generalist in nature and can be applied to a large variety of problems. A wide range of algorithms exist, and new ones are continually proposed\cite{biomimetics8030278}. This includes genetic algorithms, mathematical programming, Bayesian optimization, machine learning, reinforcement learning or other BBO methods. Most of these algorithms are heuristic in nature, making them non-deterministic. This means that two different runs, even with the same starting points, can yield different solutions. Moreover, the performance of an algorithm often vastly depends on the specific optimization problem at hand\cite{wolpert1997nofree}, making it challenging to compare different algorithms. Consequently, BBO suffers from poor reproducibility, hindering the identification of the most efficient algorithms\cite{rlgoogle}.
 Combining rigorous benchmarking\cite{gould2003cuter, gould2015cutest, dimacs, bbob, lsgo, mlda, olympus, dart, pybullet, mujoco, brockman2016openai} and BBO libraries is the only approach able to address the reproducibility crisis in optimization, ensuring transparency and reliability of results. A lot of efforts have thus been made to design benchmarks for BBO\cite{stuck, leakage, sorensenmetaheuristicsmetaphorexposed2015, nevergrad}.

{\bf From optimization to inverse design in photonics.} Optimization techniques have found applications in the improvement of photonic structures. They can also be employed in retrieving experimental parameters, as for instance commonly done in ellipsometry.

Thanks to the increase in the available computational power, numerical optimization has been increasingly used in cases where a wide range of designs can be generated. This wide range is achieved either by a low level of constraints on the geometry of the structures, or by the versatility of their optical behavior. Using numerical optimization to yield novel solutions is usually called inverse design, even though it is difficult to determine precisely when an optimization problem becomes an inverse design problem. Inverse design problems generally require advanced methods and increased computational power to solve them effectively.

For instance, problems characterized by an almost completely unconstrained geometry and often called "free form"\cite{whiting2020meta, chen2022algorithm} are considered inverse design problems beyond any doubts. However, optimizing a multilayered structure consisting of 20 layers with alternating refractive index can also be considered as an inverse design problem. In that case, even though the geometry is somewhat limited, the structure still presents a wide diversity of optical responses,  requiring advanced or adapted methods to tackle the challenge\cite{tikhonravov1996application}.

{\bf Topology optimization.} With the increase in available computational power, it has become recently possible to divide a given region of space into pixels/voxels that can typically be either filled with a material or left empty, in order to look for a structure presenting a desired optical response. Then, the number of parameters is intentionally very large, to offer the algorithms a large number of degrees of freedom. Such an approach is called Topology Optimization (TO). TO draws inspiration from its successful application in the field of mechanics, where it has been widely used. Given the large number of parameters used for representing a structure, TO usually employs gradient-based methods.  First, the problem is made continuous instead of discrete (continuous properties are allowed for a given pixel instead of a binary choice between filled and void) and since the gradient can be computed for a low computational cost (through the use of the adjoint method \cite{ceaConceptionOptimaleOu1986}), steepest descent algorithms seem a natural choice.
This approach has been extremely successful in mechanics, to the point that global optimization has been considered obsolete \cite{sigmund2011usefulness}. 

When applied to photonics problems, this approach has at first shown remarkable success, demonstrating that photonic structures can be miniaturized to an unprecedented degree while maintaining high optical performance\cite{piggott2015inverse}. However, the physics fundamentally differ in some aspects between mechanics and photonics. Mechanical problems can be regarded as comparatively simpler since a continuous deformation of an object results in a continuous and nearly monotonic deformation of its properties. The optimization process in photonics poses greater challenges compared to mechanical cases, primarily due to the wave nature of the problem. Photonic structures can exhibit resonant behavior, with each resonance corresponding to a local maximum or minimum of the cost function. 
This characteristic poses challenges for gradient descent methods, which are better suited for problems with smoother landscapes of the cost function. In photonics, two different starting points in the optimization process most often lead to two different outcomes\cite{su2020nanophotonic}. This is in contrast to mechanics, where different starting points do not typically result in significant variations in the outcomes\cite{wang2003level,sigmund2011usefulness}. Moreover, the structures produced by these algorithms often exhibit excessive intricacies, which can pose challenges for fabrication and hinder their commercialization potential\cite{molesky2018inverse, wang2018maximizing}.

{\bf Global optimization for photonics.} It is now evident that early attempts at using genetic algorithms for optimizing simple photonic structures were unsuccessful\cite{martin1995synthesis}. Specifically designed heuristics have in general failed to produce structures that are convincing enough to persuade the community to embrace global numerical optimization as a tool for years. Moreover, for continuous problems, even modern global optimization algorithms often yield unsatisfactory solutions when the parameter space has a high dimensionality, as in TO. The first successes of gradient-based TO may thus suggest that global optimization, as in mechanics, cannot compete with TO on inverse design problems. 

However, in our experience with continuous problems and modern optimization algorithms, we have found that keeping the number of parameters relatively low, typically below 100, usually suffices to give BBO algorithms enough degrees of freedom to yield innovative and convincing results\cite{barry2020evolutionary}. This approach requires limiting the number of degrees of freedom, a strategy also referred to as Parametric Optimization (PO). It becomes however challenging to make a fair comparison between BBO and TO, as they operate in different regimes.

In addition, global optimization algorithms may be valuable in problems that are typically considered in TO, due to the discretization requirement: whereas GO and BBO refer to (not incompatible) algorithm categories, TO refers to a category of problems, hence we can have simultaneously GO/BBO/TO or none of them.
Many topology optimization problems are inherently discrete, and the use of continuous parameters is primarily for leveraging gradient-based methods. The variety of the results obtained using continuous parameters\cite{su2020nanophotonic} suggests that making the problem continuous is also making it more difficult, for example because local optimization algorithms get stuck in minima with intermediate values of the refractive index.
In such cases, discrete global optimization algorithms may actually offer advantages and recent results seem to indicate that such methods are able to yield efficient solutions in a relatively consistent way, even spontaneously generating welcome features like symmetry \cite{teytaud2022discrete} and the dissapearance of intermediate refractive index values. Once again, the challenge is to identify the most suitable algorithm for a given problem, which can only be done with comprehensive algorithm libraries. In the present paper, we focus on PO, due to its advantages: suggested designs are frequently smooth and possible to manufacture, and the dimensionality typically remains moderate (dozens to hundreds).

In our opinion, global optimization algorithms are too often overlooked due to the challenges posed by high-dimensional parameter spaces. We underline that most often, finding the right algorithm makes a tremendous difference and that, given the inherent complexity of optimization, a rigorous methodology must be applied to achieve satisfactory solutions. While these methods may be computationally demanding, the advent of parallelism has made many global algorithms (which are usually intrinsically parallel) significantly more efficient, making them increasingly relevant.

\subsection{Typical photonics test cases}

\begin{figure*}
    \centering
    \includegraphics[width=\linewidth]{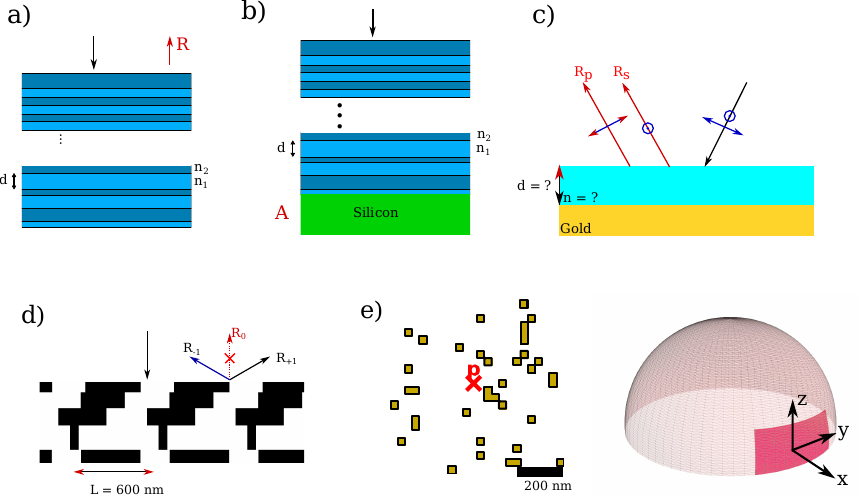}
    \caption{{\bf Graphical presentation of the test cases}.
    a) {\bf \texttt{Bragg} test case}. The objective is to maximize the reflectance of a multilayered structure composed of two alternating materials (of refractive index 1.4 and 1.7) at a wavelength of 600~nm. The parameters are the thicknesses of the different layers, but the refractive indexes are set.
    b) {\bf \texttt{Photovoltaic} test case}. The objective is to maximise the short circuit current and thus the absorption in the visible spectrum range (wavelengths from 375 to 750 nm) in the silicon layer, using an antireflective multilayered coating composed of two alternating materials.
    c) {\bf \texttt{Ellipsometry} test case}. The objective is to retrieve the thickness and the refractive index of an unknown material, using the reflectance spectrum of a single layer in both polarizations.
    d) {\bf \texttt{2D grating} problem}. The objective is to minimize the blue ($\lambda = 450$ nm) specular reflection while maximizing diffraction orders. The parameters subject to optimization are the width, position and thickness of each block of matter.
    e)  {\bf \texttt{3D nanoantenna} problem}. 
    Right : The optimization goal is to direct the emission from a local dipole lightsource towards a defined solid angle, by maximizing the ratio between the power emitted in the target direction (in red) over power emitted in the rest of the solid angle.
    Left : Top view of a geometry of a directional plasmonic gold nanoantenna. The parameters subject to the optimization are the position (in x- and y- directions) of each of the nanocubes.}
    \label{fig:all_test_cases}
\end{figure*}

We have chosen three test cases that are typical of problems that can be encountered in photonics, on which we applied the methods we recommend and benchmarked different algorithms. A Jupyter Notebook\cite{bennet_2023_10228377} is provided in which we show how cost functions can be easily defined in the context of multilayered structures using the PyMoosh module\cite{moreauPyMoosh2023} and how all the algorithms implemented in the comprehensive Nevergrad library\cite{nevergrad} can be subsequently tested. These three test cases are: a dielectric mirror optimized for high reflectivity, an ellipsometric problem, and a photovoltaic cell optimized for absorption inside the active material. These problems are represented Fig. \ref{fig:all_test_cases}. 

{\bf High reflectivity problem.} The cost function is defined as  $1-R$ 
where $R$ represents the reflectance at the working wavelength of 600~nm of a multilayered structure with alternating refractive indexes ($n_1=1.4$ and $n_2=1.8$, starting with $n_2$) as shown on Fig. \ref{fig:all_test_cases}a). Therefore, the optimization algorithms will aim to find highly reflective structures.
We considered two sub-cases (i) a structure with only 10 layers (\texttt{minibragg}), which is representative of low dimensionality problems in photonics (ii) a structure with 20 layers (\texttt{bragg}), a number we feel marks the beginning of the domain of inverse design problems.

The thickness of each layer is taken between 0 and the thickness of a half-wave layer for the lower index. When considering only a single working wavelength, adding or removing a half-wave layer to a given layer has no impact on the optical response (this is called an absent layer). Therefore, considering larger thicknesses would only introduce degenerate minima. We underline that letting the refractive index vary would lead to the same solutions, as the optimization would converge to the highest index contrast between two layers\cite{barry2020evolutionary}. This can be considered as physically expected, as this is known to be the most efficient choice to modulate the optical response of a multilayer\cite{tikhonravov1993some}.
The Bragg mirror, a periodic solution, has been identified as the best solution to this problem so far\cite{barry2020evolutionary}. It is a local optimum, and it outperforms any disordered solution, suggesting that it might be a regular, principled solution that might be the optimum -- even though it is not, as of now, strictly proven. This is why we have selected it as a test case, for which we know the (likely) global optimum, and called it ``\texttt{Bragg}''. We underline that structures reminiscent of Bragg mirrors emerge often even in 2D or 3D structures\cite{gondarenko2006spontaneous,bruleMagneticElectricPurcell2022}, for instance in waveguide problems\cite{su2020nanophotonic, teytaud2022discrete,yang2023inverse}.

{\bf Ellipsometry problem.} The objective here is to find the material and the thickness of a reference layer, knowing its reflectance properties, obtained using spectroscopic ellipsometry. This step is required to extract the desired information from ellipsometric measurements. For a tested layer with a given material and thickness, we compute the relation $e = \frac{r_p}{r_s}$ where $r_p$ is the reflectance of the whole structure in TE polarization and $r_s$ is the reflectance in TM polarization, for a wavelength range of 400~-~800~nm and for a given incidence angle of 40\textdegree. The quantity we minimize here is the difference between the $e$ computed from the tested layer and the $e$ computed from the reference layer. 

The thickness of the tested layer is taken between 30 and 250~nm, and we assume the refractive index is comprised between 1.1 and 3. This simple problem is illustrated in Fig.\ref{fig:all_test_cases}c). Problems in ellipsometry are generally more complicated but highly dependent on the response model assumed for the material. Our simple dispersion-less and absorption-less test case can however be adapted easily thanks to the architecture of PyMoosh\cite{pymoosh}.

We underline that in many ways, this case mirrors the practical challenges faced by photonics researchers. It illustrates the common situation where researchers design structures characterized by a small number of parameters, and then engage in optimization to determine the upper limits of their performance\cite{smaali2021reshaping,centeno2021inverse,moreau2007enhanced}. Such a problem is typically not considered as inverse design.

{\bf Photovoltaics problem.} The objective here is to find the best possible antireflective coating in normal incidence with a multilayer structure made of two alternating materials (permittivities of 2 and 3 respectively) on an amorphous hydrogenated silicon substrate (30000 nm).   The quantity to be maximized here is the short circuit current in the 375~-~750~nm range, assuming a quantum yield equal to 1, as described in \cite{barry2020evolutionary,bennet2021analysis}. 

The cost function $f$ is one minus the efficiency defined as the ratio between the short circuit current $j_{sc}$ and the maximum achievable current $j_{max}$ if all photons are converted into electron-hole pairs :
\begin{equation}
f=1-\frac{j_{sc}}{j_{max}}
\end{equation}
with
\begin{equation}\label{eq:jsc}
j_{sc} = \int A(\lambda) \frac{d I}{d\lambda}.
 \frac{\mbox{e}\lambda}{\mbox{h} \mbox{c}}\,\mbox{d}\lambda,
\end{equation}
where $A(\lambda)$ is the absorbance of the active layer, e, h and c are respectively the elementary charge, the Planck constant and the speed of light in vacuum and the spectral density of the illumination $\frac{d I}{d\lambda}$ is given by the solar spectrum\cite{santbergen2010am1}.

The problem is illustrated in Fig.~\ref{fig:all_test_cases}b).
As for the high reflectivity problem  we considered three sub-cases (i) a structure with only 10 layers (\texttt{photovoltaics}), (ii) a structure with 20 layers (\texttt{bigphotovoltaics}), and (iii) a structure with 32 layers (\texttt{hugephotovoltaics}). 

\subsection{Complex photonics cases}

In order to show how the techniques we advocate for can be applied to much more complex cases, we have selected 2D and 3D cases that can be studied even though they require much more computational power compared to multilayered structures.

The 2D case corresponds to a grating  with a 600 nm period and composed of 5 layers containing a single block of matter, whose size and position can be adjusted as illustrated in Figure \mbox{\ref{fig:all_test_cases}d)}. The cost function aims to minimize the specular reflection and maximize the diffraction orders, producing structures resembling the photonic structures which can be found on the wings of Morpho butterflies.

The 3D problem focuses on the optimization of a directional plasmonic nanoantenna that couples with a local quantum emitter, and steers the light towards a defined solid angle, as illustrated on the Figure \mbox{\ref{fig:all_test_cases}e).}

For more details, the interested reader is invited to consult our previous work\mbox{\cite{wiechaDesignPlasmonicDirectional2019, barry2020evolutionary}} and corresponding codes\mbox{\cite{bennet_2023_10228377}}. A first notebook shows how to perform an optimization using DE for the grating case, which generally produces a regular structure. Two notebooks are devoted to the plasmonic antenna -- one showing a simple optimization and the second one leveraging the parallel library MPI to benchmark algorithms on that particularly costly test case.

\section{Basic tools for optimization: algorithms and observables} 

As discussed before, many algorithms and methods are available to perform optimization and eventually inverse design. It is therefore necessary to define some observables and basic tools to compare the performances of algorithms in a reliable way. This section presents first some well known categories of algorithms. It is shown how these algorithms can all be run through the Nevergrad platform\cite{nevergrad}. Then, we define relevant 
observables, \textit{i.e.} quantities allowing discussions about the performances of algorithms and their results. All the discussions in this section are illustrated by results of the codes provided in the supplementary material\cite{bennet_2023_10228377}.

\subsection{Algorithms categories}

Many BBO platforms exist, including CMA~(Covariance Adaptation Algorithm)\cite{pycma}, AX~(Adaptive eXperiments)\cite{ax}, HyperOpt~\cite{hyperopt}, SMAC (Sequential Model-based Algorithm Configuration~\cite{smac}), NLOPT\cite{nlopt} and Nevergrad\cite{nevergrad}: Nevergrad is a free Python library which imports all those ones and others. We present in this section the algorithms from Nevergrad used in our benchmark and in our Jupyter Notebook experiments\cite{bennet_2023_10228377}, and organize them based on their respective categories. 

{\bf{Mathematical programming.}} \label{section:mathprog}Methods available for gradient-based optimization have been adapted to the black-box setting.
For example the limited-memory Broyden-Fletcher-Goldfarb-Shanno LBFGS~\cite{lbfgs} method can be applied, with derivatives computed by finite differences. We underline that other methods for computing the gradient more efficiently are available, especially the adjoint method\cite{allaire2015review} widely used in TO, or numerical tools\cite{maclaurin2015autograd}, but this is beyond the scope of the present paper.

{\bf{Genetic and Evolutionary Algorithms.}}
One of the most well known families of algorithms are evolutionary and/or genetic algorithms.
In an evolutionary algorithm, each step of the process creates and evaluates the cost function of a ``population'' of structures, preserving the ``individuals'' (\textit{i.e.,} the structures) that are better than those belonging to the population of the previous step (also called the previous ``generation'').
The different genetic algorithms have different ways of creating the new generation of structures. Most include steps such as mutations and crossovers between structure information.

We underline that the historical algorithms, in which for instance the binary coding of the parameters is considered as its genetic code, are considered obsolete in the optimization community because of their well documented lack of efficiency\cite{janikow1991experimental,herrera1998tackling} and their use should be avoided.

A lot of more efficient algorithms, inspired by these early ideas, have emerged over the years, with an improved efficiency. One of the most well known and efficient overall seems to be Differential Evolution (DE~\cite{de}), an algorithm that has many variants. One of the most classical variants is presented in Table~\ref{algo:depseudo} and the different formulas that can be used for the mutation to define new variants are presented in Table \ref{algo:demutations}. By default, we use the current-to-best version of DE. 

In our experiments, we often use the quasi-oppositional variant QODE \cite{quasiopposite}, which randomly draws half the initial population, and then, for each point $x$, adds $-r\times x$ to the population with $r$ randomly uniformly drawn in $[0,1]$ (assuming that the center is $0$: otherwise $c-r\times(x-c)$ with $c$ the center of the optimization domain). We also include QNDE (Quasi-Newton DE), which runs QODE for half the budget and then finishes with BFGS with finite differences: this is a memetic algorithm, which runs first a global algorithm and then a local one.

\begin{table}[h]
\begin{algorithmic}
  \STATE{Randomly draw the population $x_1$,\dots,$x_{30}$ in the search space.}
  \WHILE{There is time}
     \STATE{\COMMENT{$//$ This is one iteration}}
     \FOR{each $x$ in the population}
        \STATE{Randomly draw $a$, $b$, $c$ and $d$ individuals (i.e. designs)} in the population.
        \STATE{Compute $y$ using one of the equations in Table \ref{algo:demutations}.\COMMENT{$//$Mutation operator}}
        \STATE{\COMMENT{$//$Crossover operator: for loop below.}}
        \STATE{Randomly draw $i_0$ in $\{1,\dots,d\}$.}\ \ \ \ \  \COMMENT{$//$ $d$ is the dimension}
        \FOR{each $i$ in $\{1,\dots,d\}$}
             \IF {{$i=i_0$} or with probability $CR$}
             	\STATE{$z_i\leftarrow y_i$.}
             \ELSE
             	\STATE{$z_i\leftarrow x_i$.}
             \ENDIF
        	 \IF{$z$ has lower cost function value than $x$}
             	\STATE{$x$ is replaced by $z$ in the population.\COMMENT{End of crossover operator}}
        	 \ENDIF
        \ENDFOR
     \ENDFOR
  \ENDWHILE
\end{algorithmic}
\caption{\label{algo:depseudo}{\bf Pseudo-code of DE}. $a,b,c,d$ are distinct, randomly drawn individuals in the population and $best$ refers to the best design so far. The $y$ individual is obtained by a mutation formula (see Tab.~\ref{algo:demutations}) and $CR$ is the crossover rate, indicating the percentage of the mutant $y$ used in the creation of a new individual $z$. {$i_0$ ensures that at least one variable is mutated.}}
\end{table}

\begin{table}
\begin{eqnarray*}
\mbox{DE/rand/1:\ \ \ }             & y(x) = & a + F_1(b-c) \\
\mbox{DE/best/1:\ \ \ }             & y(x) = & best + F_1(a-b) \\
\mbox{DE/randToBest/1:\ \ \ }       & y(x) = & c + F_1(a-b) + F_2(best-c)\\
\mbox{DE/currToBest/1:\ \ \ }       & y(x) = & x + F_1(a-b) + F_2(best-x)\\
\mbox{DE/rand/2:\ \ \ }             & y(x) = & a + F_1(a-b+c-d) \\
\mbox{DE/best/2:\ \ \ }             & y(x) = & best + F1(a-b+c-d)
\end{eqnarray*}        
\caption{\label{algo:demutations} {\bf Various DE formulas}. $a,b,c,d$ are distinct, randomly drawn individuals in the population {and are thus vectors containing all the parameters describing the structure}.  We see that the mutated variant of $x$, namely $y(x)$, is in some cases, by design, independent of $x$. $best$ refers to the best design so far. Typically, $F_1=F_2=0.8$ and $CR=\frac12$. DE is called differential because it is based on the difference between vectors. We underline that, if all the vectors of the population have the close values for a parameter and since DE is based on differences, this parameter will not change much generation after generation. Efficient sub-parts of the vectors are thus preserved during the optimization.
} 
\end{table}

All the variants of DE typically perform well for a large parameter space dimension , including quite irregular cost functions such as the ones which appear in photonics. Most winning algorithms for Large-Scale Global Optimization (LSGO) competitions use a variant of DE\cite{lsgo}. 

There are many reasons why DE can be a sensible default choice - among them its simplicity, its low computational overhead, and its good performance overall {associated to a general robustness to changes in the optimization conditions}. We also point out that the rise of parallel computing resources makes DE and other global methods like PSO {(see below) faster}: whereas parallelizing BFGS or other gradient-based methods is tricky, DE (as implemented and parametrized in the present paper and in most implementations) is just 30 times faster with 30 cores than on a sequential machine, and can be {even more} parallel with a simple adaptation of parameters (typically increasing the population size). Besides this natural parallelism advantage, many black-box optimization methods can include arbitrary parameters (including discrete parameters, e.g. choosing a material in a list), adding non-differentiable stochastic {manufacturing errors\cite{robustoptim,robustoptim2},} adding multiple objective functions and handling worst-case analysis over input angle or over a range of wavelength.

As its name indicates, DE is built on a differential formula to create new structures based on existing ones. This means that if multiple structures in the current population share the same characteristics (e.g., in the present framework, the same block of layers with identical thicknesses and materials), those characteristics will probably be preserved in the creation of the new candidate solution. In photonics, this leads to the emergence of modular structures with distinct blocks of layers, each having well-defined optical functions. This property might contribute to its efficiency addressing photonic problems~\cite{barry2020evolutionary,rapin2020open}. Also, DE can deal with discrete spaces as well, e.g. choosing between many materials.

Other evolutionary methods include Particle Swarm Optimization (PSO~\mbox{\cite{pso}}): a population of particles with random initial positions and velocities is attracted towards good search points found so far. PSO is known as a robust method, dedicated to rugged cost functions.

Typically, Evolution Strategies (ES) iteratively update a probability distribution model, that is used to optimized the likelihood for generation of good candidates. A well-known example is Covariance Matrix Adaptation (CMA~\cite{CMA}) which updates the entire covariance matrix of a multivariate normal distribution (MND). The CMA algorithm samples candidates using a MND, evaluates them, and then modifies the MND using the evaluation results.

While CMA typically performs best on the Black Box Optimization Benchmark (BBOB\cite{bbob}), simpler methods such as the $(1+1)$ evolution strategy with one-fifth rule\cite{rechenberg73} are still good in many cases due to its ability to quickly adapt the search scale. 

{\bf{Bayesian Optimization.}}
In a Bayesian optimization process\cite{ego,elsawy2021multiobjective}, the algorithm uses a probability distribution (a Gaussian process) for modeling the cost function and the uncertainty, and updates it with each new structure tested. The model provides, for each point in the search space, an estimated probability distribution for the cost function value of that search point.

Therefore, it is possible, at each iteration, to define the {\em{Expected Improvement}} (EI) for a search point $x$: 
$EI(x) = \mathbb{E}_w  \max(0, m - CostModel(x,w))$, 
where $m$ is the best cost so far and $CostModel(x,w)$ is the current statistical model of the cost function. $CostModel(x,w)$ depends on a random $w$, because this model is statistical.
The value of $w\mapsto CostModel(x,w)$ is updated each time a new cost value is available: it is frequently a Gaussian process.

This gives an estimation of what structure to try next: we search for $x$ in the search space such that $EI(x)$ is approximately maximal -- this requires a side optimization for each iteration, hopefully orders of magnitude faster than the original problem. 

Many variants of this approach have been proposed in the literature, with many parameters for the stochastic model. By design, Bayesian optimization (BO) works best with problems with a small parameter space dimension and for relatively smooth functions or functions adapted to the design of the kernel used in a specific BO implementation. Also, the probabilistic interpolation process, which is necessary to know which structure to try next, can be expensive, possibly becoming computationally more expensive than the cost function itself.For these reasons, BO is best suited for problems with a cost function that is computationally costly\cite{elsawy2020numerical}. In such cases, BO is then difficult to compare to other approaches, which explains why comparative studies are not abundant. For instance, the authors in \cite{bbobsmac} tested Bayesian Optimization in a limited budget scenario but other, computationally cheaper algorithms were still performing well. As BO are not relevant in the context of our test cases, no BO method has been benchmarked in the present work, but for computationally expensive cost functions, such methods are often preferred.

{\bf{Other black-box optimization methods.}}
Other methods include pattern search methods, such as e.g. the Nelder-Mead method\cite{NM}, which iteratively updates a simplex of search points.
Also, optimization wizards (also known as hyper-Heuristics) are defined by automatically selecting and combining existing methods\cite{versatile}: NGOpt (Nevergrad optimizer) and NGOptRW (NGOpt real-world) are examples of wizards included in Nevergrad. These are home-made Nevergrad algorithms, combining many algorithms (including DE and BFGS) with selectors tuned on a wide range of methods.
Methods based on reinforcement learning (RL) have also been defined and applied on machine learning problems\cite{NAS}, though simple classical evolutionary methods were actually competitive~\cite{evo}. Please note that a study\cite{omcd} mentions difficulties for reproducing results in some RL papers.

\subsection{Observables}
\label{section:observables}
For the vast majority of continuous optimization problems, proving that a solution provided by an algorithm is the best possible solution in the optimization domain is essentially impossible. In addition, many optimization algorithms are non-deterministic or sensitive to the initialization, which means each run of the algorithm will lead to a different outcome. 

As a consequence, it is possible to perform many optimization runs and still not be able to firmly determine whether the best of these solutions is good enough to stop looking for other, potentially better, outcomes. Yet, observing how exactly each different run progresses towards a solution, as well as considering the solutions that are produced {\em statistically}, yields crucial information that may help the user gain confidence in the best solutions produced or, on the contrary, indicate that these solutions cannot be trusted. 

{\bf Convergence curves}. The first observable, which is widely used, is the convergence curve that can be produced for each run of a given algorithm by plotting the value of the cost function of the solution that the algorithm {\em recommends} (typically the best solution found so far) as a function of the number of iterations. When multiple runs are launched, convergence curves can be either drawn for each run or an averaged convergence can be obtained. Both essentially provide the same kind of information: whether most of the runs have settled on a given solution (if they have almost all reached a plateau for many iterations), or if further iterations have a chance to improve the outcome. 
Fig. \ref{fig:convergencecurve_1} (resp. Fig. \ref{fig:convergencecurve_2}) presents such an example of individual curves for each run (resp. aggregated curves with average convergence).

{\bf Budget}. Not all iterations of all algorithms have the same computational cost or evaluate the cost function the same number of times. An iteration for a genetic or evolutionary algorithm typically corresponds to a new generation and thus the evaluation of a few dozen of solutions. For some other algorithms, an iteration requires the evaluation of a single new solution. A way to compare two algorithms fairly is to discuss their performances for a given number of evaluations of the cost function. The maximal number of evaluations allowed is called the budget and it has to be chosen before running the algorithms. Each run of an algorithm is then stopped when the budget has been reached. Of course, this does not take into account the computational cost of the optimization algorithm itself, which should be discussed when not negligible compared to the cost function values.

{\bf Consistency curve}. Convergence curves for different runs allow to determine whether the chosen budget is large enough to reach a satisfactory convergence for each run. However, since the different runs can produce completely different results, the variability of the different results has to be discussed. This can be done by plotting what we call the consistency curve (Fig. \ref{fig:consistencycurve}). This curve is obtained by plotting the different values of the cost function reached at the end of each run, sorted from the lowest to the highest. This is generally done for the highest budget allowed. When such a curve displays a plateau, then the same solution, or at least solutions with the same cost, has been found several times by the algorithm. A large plateau for the lowest value of the cost function thus means the best solution has been found a large number of times. This reinforces the confidence that can be placed in that solution.

{\bf Box plots}. It is not always relevant to draw the consistency curve, especially when comparing a large number of algorithms. This curve can be summarized by a box plot as \textit{e.g.} in Fig.~\ref{fig:boxplots}. While a box plot will not allow to observe any plateau and thus bears less information than a consistency curve, it gives an immediate and easily readable access to the statistical properties of the values of the cost function. 

{\bf Estimating the density of local minima}. In addition, local optimization algorithms can be used to estimate whether there is a limited or a large number of local minima and whether the best solution found by any other algorithm has a large attraction basin or not. A simple way to do so is to launch multiple runs of the algorithm with starting points drawn randomly in the optimization domain, then make sure all the runs have converged (e.g. BFGS has a stop criterion essentially guaranteeing a local minimum has been reached) and finally plot the resulting consistency curve. If we were to run a large number of such optimizations, the consistency curve should present plateaus corresponding to different local minima  -- several minima may present identical values of the cost function, due to symmetries of the problem, and not be distinguishable. The width of a plateau would allow to estimate the volume of the corresponding attraction basin.

In this context, running BFGS with randomly drawn starting points can be seen as a way of estimating the difficulty of a problem. Many different results suggest a lot of different minima and if the best result is rarely found, this also means it has a relatively small attraction basin compared to the size of the optimization domain. These characteristics are indicative of a difficult problem.

\section{Running and discussing experiments: the art of benchmarking}

In this section, after presenting our codes and explaining how to reproduce the results below, we evaluate selected algorithms using our multi-layer test cases. We provide criteria for selecting the most efficient algorithms. In addition, we carry out an in-depth physical analysis of our solutions to assess their quality. 

\subsection{Repository description}

With the idea that anyone should be able to build easily on our work, we provide Jupyter Notebooks and Python codes in a repository\mbox{\cite{bennet_2023_10228377}} to demonstrate how to perform global optimizations of multi-layered photonic structures, 2D gratings and 3D plasmonic nanostructures.

The cost function computation is necessarily based on Maxwell's equations solvers, which are adapted to the geometry considered. Multi-layer simulations are done with PyMoosh, a Python-based simulation library designed to provide a comprehensive set of numerical tools allowing to compute essentially all optical characteristics of multilayered structures and that we released recently\mbox{\cite{langevin2023pymoosh,moreauPyMoosh2023}}. The optical response of the 2D gratings is obtained using a home-made Rigorous Coupled Wave Analysis (RCWA) code\mbox{\cite{lalanne1996highly,granet1996efficient}} whereas the plasmonic nanostructures' scattering diagram is computed with pyGDM, a freely available Green Dyadic Method Python implementation\mbox{\cite{wiecha_pygdm_2018, wiechaPyGDMNewFunctionalities2022}}. As optimization toolkit we use either Nevergrad\mbox{\cite{nevergrad}}, especially for benchmarking and sometimes PyMoosh's internal DE optimizer when this is sufficient.

Our repository contains simple notebooks to show, on the \texttt{Bragg} problem, (i) how to use PyMoosh's internal DE optimizer, (ii) how to use a Nevergrad optimizer, and (iii) how to leverage Nevergrad to benchmark algorithms. More notebooks are provided to showcase an optimization using Nevergrad (iv) on the \texttt{Ellipsometry} test case, (v) on the \texttt{Photovoltaics} test case, (vi) on the 2D grating problem, and (vii) on the 3D nanoantenna problem.

Then, more complete examples are also presented, that will require supplementary steps as well as more computational power to run. The first example, in a separated folder, is a Nevergrad-based benchmark of several algorithms on the 3D plasmonic nanoantenna test case, accelerated by relying on MPI for the parallelization. The second folder contains codes allowing to reproduce all the benchmarking results presented in this paper -- including, importantly, the convergence and consistency curves on all multi-layer test cases. We even provide a simplified notebook that can be run directly on the Colab platform, in order to maximize the reproducibility of our results and respect the principles of the FAIR approach (making data and methods Findable, Accessible, Interoperable, and Reusable).

The first examples, presented in notebooks, are voluntarily kept simple to be run on any platform, and we underline that they do not always present the same optimization configuration (in terms of budget, algorithms or initialization) as the comprehensive code. As a consequence, the Notebooks may be produce results that differ from the benchmark presented below.

\subsection{Benchmarking with Nevergrad}

We present and discuss here benchmark results that can be obtained using the codes described above. We focus here on the methodology and on the information that can be obtained by closely monitoring the observables we have defined earlier. We have thus limited the number of algorithms for which we show optimization results.

\begin{figure}\center
\includegraphics[width=\linewidth]{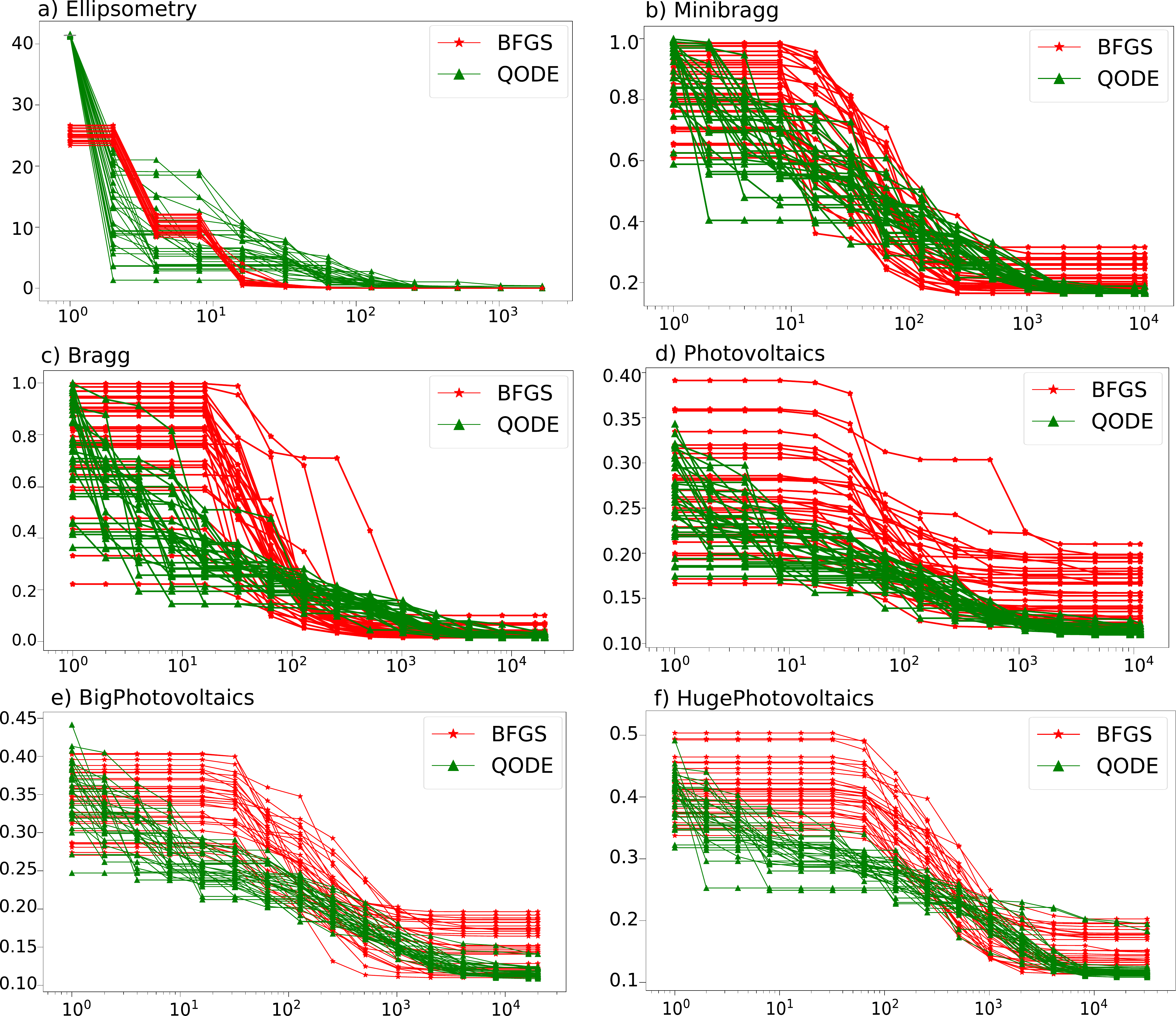}
\caption{\label{fig:convergencecurve_1} {\bf Convergence curves}. Convergence curves for different runs for different algorithms showing the value of the cost function as a function of the number of iterations performed so far by the optimization. The lower the algorithm is in the legend, the better its performances are. The information brought by such individual curves is useful but difficult to read when comparing even a small number of algorithms, that is why the average convergence is often plotted, as shown in Fig. \ref{fig:convergencecurve_2}. }
\end{figure}

\begin{figure}\center
\includegraphics[width=\linewidth]{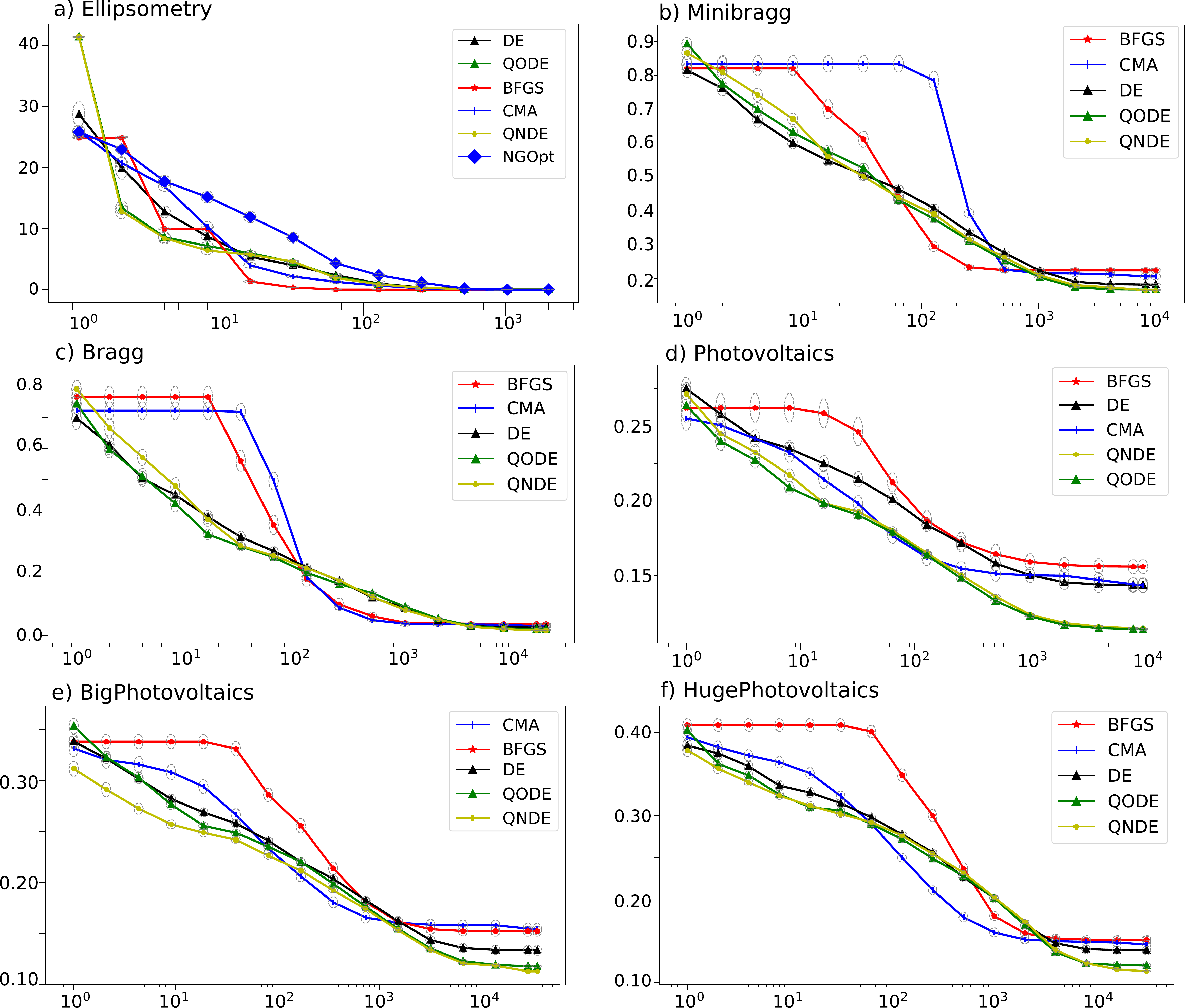}
\caption{\label{fig:convergencecurve_2}{\bf Averaged convergence curves}. Averaged convergence curves for different algorithms (31 runs each), with standard error visible in ellipses, as a function of the number of iterations.  The algorithms are ordered by average performance for the maximum budget (the lower in the legend the better). Using an average convergence curves allows to compare more algorithms\cite{schneider2019benchmarking}.
}
\end{figure}

{\bf Algorithm comparison}. The convergence curves (shown in Fig. \ref{fig:convergencecurve_1}) allow to assess whether the budget is high enough. The averaged curves shown in Fig. \ref{fig:convergencecurve_2} are usually the most readable and informative, especially if the standard error is also given on each averaged curve.

\begin{figure*}\center
\includegraphics[width=\linewidth]{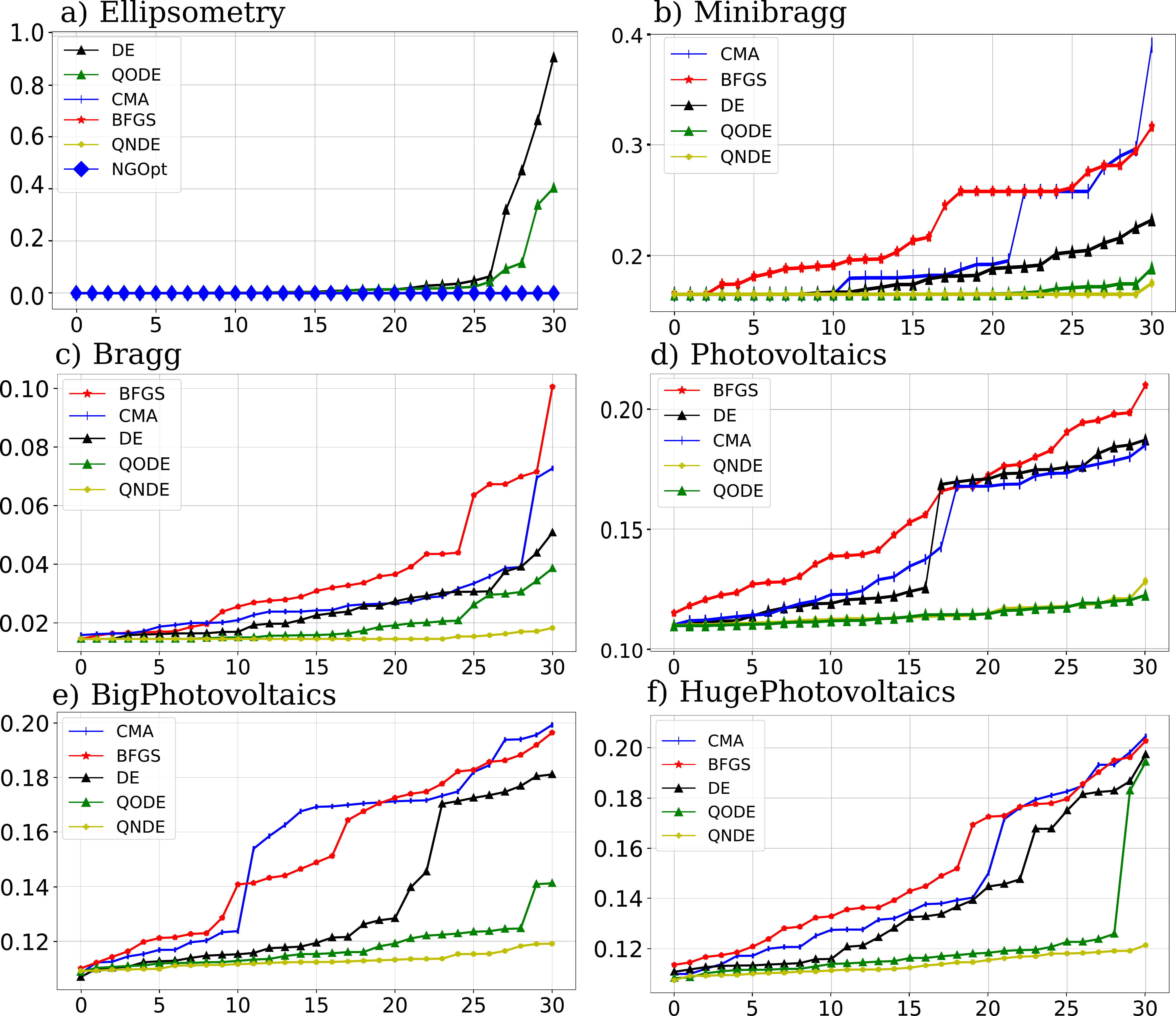}
\caption{\label{fig:consistencycurve} {\bf Consistency curves for different algorithms}. Each algorithm has been run 31 times with the highest budget (from $10^3$ to 32000 iterations depending on the test case, as shown in the convergence curves Fig. \ref{fig:convergencecurve_1}). The cost function values for each run's best solution are sorted from left to right in ascending order. The lower the curve and the flatter, the more efficient and reliable the algorithm can be considered. A plateau means that a certain cost function value, and probably the same solution, has been found multiple times, indicating a good reproducibility of the optimization. The results for BFGS and its variants correspond to different local minima if the algorithm has converged (this is often the case when the budget is large).The lower the algorithm is in the legend, the better its performances are.}
\end{figure*}

\begin{figure}
\centering
\includegraphics[width=\linewidth]{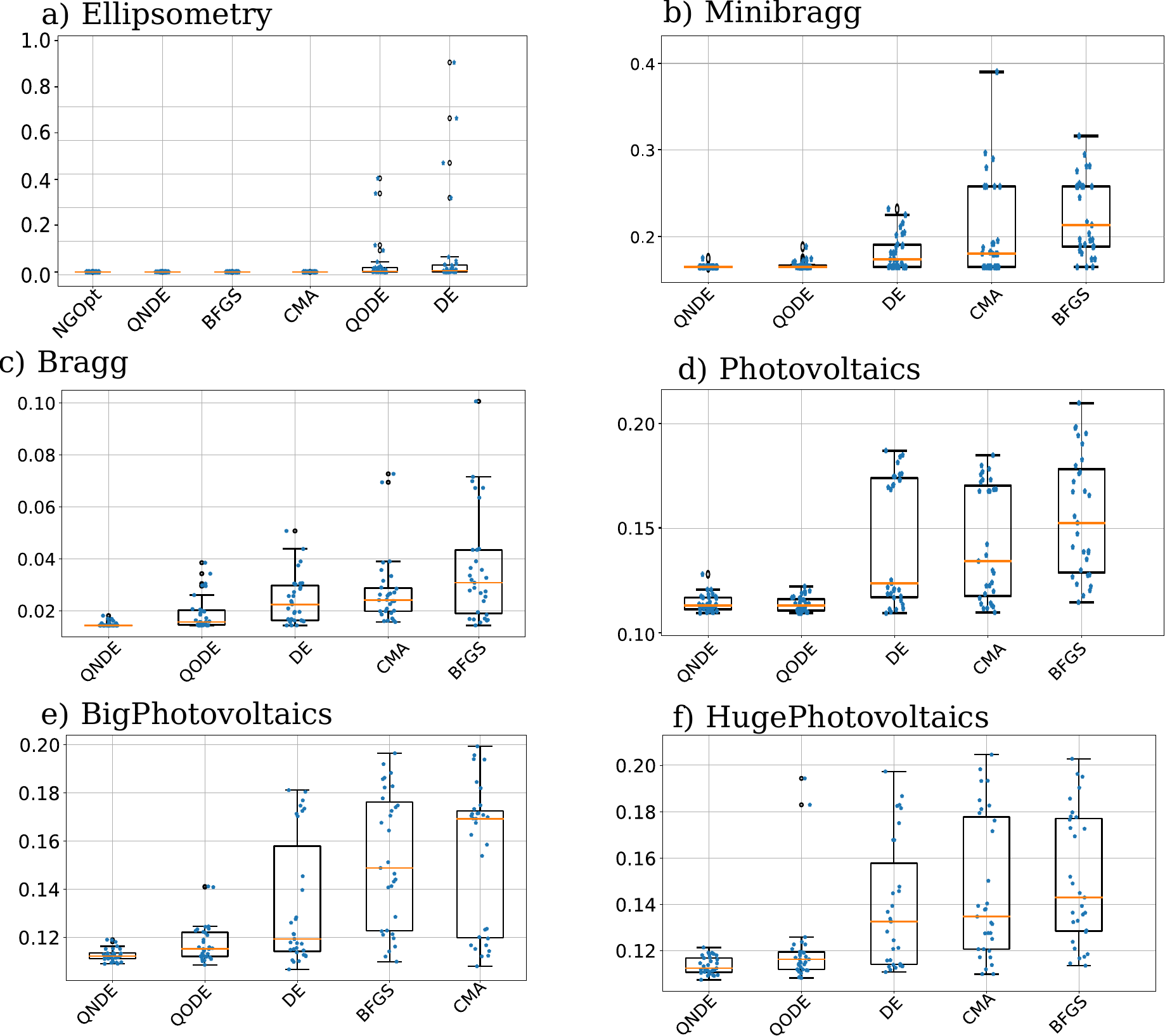}
\caption{\label{fig:boxplots} {\bf Performance box-plots for comparing algorithms}. Box-plots representing the distribution of the minimum values of the cost function reached at the end of each run for a given algorithm for different test cases. The results are shown for the highest budget, ranging from 1000 for \texttt{Ellipsometry} to 32000 for \texttt{HugePhotovoltaics}, as shown in Fig. \ref{fig:convergencecurve_1}. Each box presents the first quartile, the third quartile, a line at the median, and dashed lines are removed at 1.5 times the inter-quartile range. The algorithms are sorted according to their performances, the best performing is placed on the left.}
\end{figure}

Once convergence has been reached for all algorithms, the consistency curves shown in Fig. \ref{fig:consistencycurve} allow to compare thoroughly the reliability of different algorithms. Showing on the same graph the consistency curves for a large variety of algorithms is not convenient, as figures can be difficult to interpret. Each consistency curve can however be summarized using a box plot, allowing for a fair benchmarking between various algorithms, as shown in Fig. \ref{fig:boxplots}.

According to these results, with the optimization domain and initialization we have chosen (see below), DE and the variants we considered (QODE and QNDE) performs better than CMA on all our test cases. We underline that we have actually tested more algorithms, without showing the results (they can be obtained using the benchmark codes provided) but that CMA and DE appeared as the best options overall. CMA's performances are actually relatively close to the performances of DE with a random initialization. However, we observed that CMA often has poorer consistency.

{\bf Density of local minima}. We have also run BFGS with an almost unlimited budget, using randomly chosen starting points in the optimization domain. All the points of the corresponding consistency curve are thus actual local minima, which allows to assess their density and the difficulty of the optimization problem.
In the ellipsometry case, BFGS almost always finds the minimum, which means the density of local minima is extremely low. This is not the case for DE, for instance. It is not surprizing to see that many algorithms are efficient in that case.
Retrieving the Bragg mirror as a solution is more complex because of a relatively large density of local minima that seem to increase with the number of parameters of the problem.
Photovoltaic problems seem to be the most difficult for all algorithms, to the point that for \texttt{BigPhotovoltaics}, the performances of CMA are similar to those of BFGS, which means it does not bring an advantage compared to a steepest descent with a random start.
For 32 layers, the original DE algorithm finds a minimum close to the best only one time out of three, while CMA and BFGS find a continuum of values for the cost function, reminiscent of what happens for BFGS in complex TO cases\cite{su2020nanophotonic}.

{\bf Impact of initialization}. The bounds we have defined for the optimization domain on a physical basis have actually two roles. One is ensuring the realism of the produced solution, by forbidding nonsensical values for the parameters, such as a negative thickness for a material layer. The other role is to give the algorithms an indication of the domain in which a solution can be expected.

Often in photonics, once the bounds have been defined, algorithms are initialized by drawing the first structure randomly according to a uniform random distribution inside the optimization domain. This choice is natural for BFGS, for instance, as it allows to estimate more accurately the density of local minima in the optimization domain.
When an initial population is required, for DE or CMA for instance, we have partially used Nevergrad's way of initializing algorithms, in a way that is consistent with the standards of the optimization community. The population is generated around a center (drawn randomly according to a uniform random distribution in the optimization domain) following a normal distribution with width of a fraction of the optimization domain (typically 1/6, but it varies for certain algorithms).
Such an initialization allows one to estimate how good an algorithm is at finding a solution even if it is located outside its initialization region. This kind of initialization could also be used to target a specific region of the optimization domain, a role which is generally devoted to the bounds of the optimization domain.

The importance of initialization is well illustrated by the performances of QODE when compared to DE with the initialization described above. In QODE, for each individual randomly chosen in the optimization domain, another one is added symmetrically with respect to the center of the domain. This simple trick improves the performances of DE in all cases, which is why we recommend it. As shown by the different formulas of DE variants, DE is more efficient to explore the domain when difference between individuals is large. This initialization ensures that there will be a spread in the values chosen for all parameters in the initial population, making DE measurably more efficient. QODE has performed impressively on all our test cases.

{\bf Combining algorithms}. We underline here that combining algorithms can be a good idea if the algorithms are complementary. The memetic QNDE combines QODE and BFGS. DE provides excellent starting points for BFGS, which is then efficient at finding the closest local minimum, making such a strategy effective in all our test cases. We have also tested NGOpt, the "wizard" of Nevergrad\cite{meunier2021black}, on the ellipsometry case. It is automatically choosing a well working algorithm (note that many algorithms perform well here).

Analyzing physically both the solutions produced and how they have been produced (by which algorithm, in which conditions and how fast) sheds a new light both on the solution and on the optimization process itself. When a problem is grounded in physics, one can and should take advantage of its distinct characteristics in order to gain deeper insights.

{\bf Low dimension.} In the \texttt{Ellipsometry} case, local algorithms perform perfectly and find the solution reliably and quickly. This is typical for a simple landscape with few local minima. Physically, this means the geometry is simple enough and the considered layers sufficiently thin to have a low number of resonances. This may not always be the case in ellipsometry problems, for instance if the material dispersion is better described using a large number of parameters\cite{rakic1998optical}.

{\bf Multiple resonances and local minima}. The \texttt{Bragg} test case is more interesting. Even for ten layers, local algorithms most often fail to find solutions performing as well as the Bragg mirror. Since the algorithms have converged and since local algorithms converge to local minima, this means local minima are numerous, even for such a relatively simple problem with an intermediate number of parameters. This is expected, since for only a dozen of layers the thickness of the whole structure is already more than 2 µm, so that a large number of resonances, leading to as many local minima, can be expected in such a system. 

For photovoltaic cases, the problem is made even more complicated because of the use of a realistic (and thus noisy) solar spectrum in the cost function, which leads to even more local minima. This naturally leads to a decrease in the performance of the algorithms, as shown by the consistency curves that are less flat than for more simple cases. 

\begin{figure*}
    \centering
    \includegraphics[width=\linewidth]{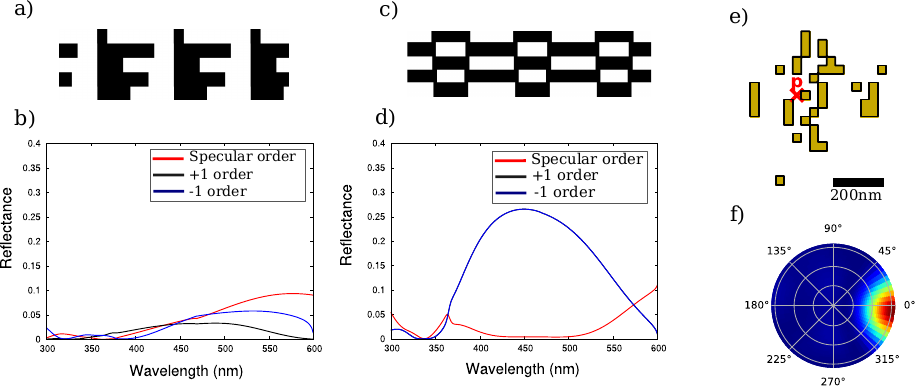}
    \caption{{\bf Optimization of 2D or 3D structures}.
    {\bf a)} Geometry of a 2D grating corresponding to a typical starting point. The dielectric material is represented in black and air in white. {\bf b)} Diffraction efficiencies in reflection for the grating represented in a).
    {\bf c)} High-performance result of an optimization for the 2D grating. 
    {\bf d)} Diffraction efficiencies in reflection for the grating represented in c). The efficiency is maximized in the $\pm 1$ reflected orders (blue and black solid lines), while it is minimized in the specular reflection (red solid line) at the working wavelength ($\lambda = 450\text{ nm}$).
    {\bf e)} Top view of an optimized geometry of a directional plasmonic gold nanoantenna, coupled to a local emitter, after multiple runs using several algorithms.
    {\bf f)} Far-field emission pattern of the optimized antenna showing the directionality of the emission.}
    \label{fig:2D_3D_problems}
\end{figure*}

For the 3D optimization case, we have run the optimization using a selection of several optimizers each with a budget of 10000 evaluations and repeating every optimizer run 10 times. The results are depicted in Figure~\ref{fig:2D_3D_problems}e, where the best structure found is shown. The corresponding emission pattern is shown in Fig.~\ref{fig:2D_3D_problems}f. The result is consistent with our former results \cite{wiechaDesignPlasmonicDirectional2019}. 

Using convergence and consistency plots (not shown but retrievable with the provided 3D codes), we find that on this problem, significant performance differences occur depending on the optimizer. We attribute this to the discrete character of the problem, which is not ideal for many continuous algorithms.  We find in particular, that gradient based optimization (BFGS) is totally unsuited for this optimization problem with discrete positional parameters. Also, the consistency curves in this problem do not show a plateau, for neither of the algorithms, indicating that the budget is not sufficiently high for convergence. The high variance of the individual runs finally indicates that probably a high number of local minima exists.

{\bf Regularity and modularity}. As explained for DE in section~\ref{section:mathprog}, if the same values for some parameters can be found in all the individuals they will be preserved throughout the optimization. This is well shown in Table~\mbox{\ref{algo:demutations}}: DE is based on the difference between vectors in all the different versions of the mutation formula, so that if a subset of parameters is present in all the individuals, these parameters evolve only slightly. In photonic structures, it is common to find structures where different parts play well definite roles (such as an anti-reflective coating typically, or a photonic crystal core). Such structures are said to be modular. DE is well equipped to find modular structures because it will tend to keep a subset of parameters (which, depending on the parametrization, can correspond to a sub-part) if that makes all the solutions more efficient.

\begin{figure}
    \centering
    \includegraphics[width=1\linewidth]{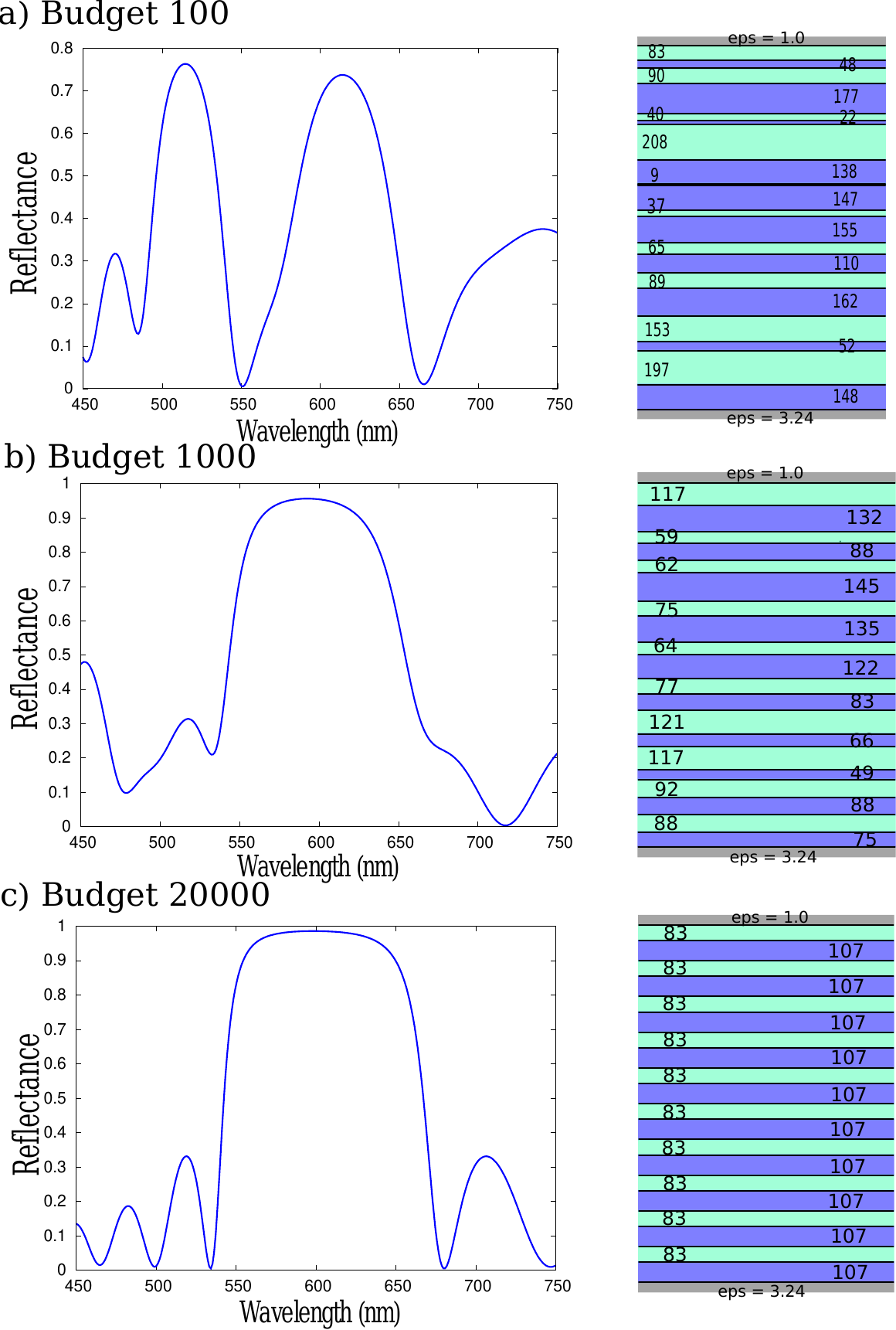}
    \caption{Best structures (right) and their associated reflectance spectrum (left) obtained for the \texttt{Bragg} case with 20 layers for a budget of {\bf a)} 100   {\bf b)} 1000   {\bf c)} 20000. The light, resp. dark color represents the high (1.8), resp. low (1.4) refractive index material. The grey color represents substrate and superstrate. 
}
    \label{fig:bragg_results}
\end{figure}

This is illustrated for the \texttt{Bragg} case in Fig. \ref{fig:bragg_results}, where the best results (whatever the algorithm) are shown for different budgets. For the lower budget, the structure appears disordered (no algorithm has converged at that point) and the spectrum indicates that a relatively narrow anti-resonance producing a large reflectance has been found. This does not qualify as a local minimum: a minor adjustment in the spectrum can align the reflectance peak at 600 nm, achievable through a straightforward contraction of the entire structure since the materials considered are not dispersive. The absence of such alignment serves as a clear indicator that the optimization falls short here.

 For a budget around 1000, the best solution, shown in Fig. \ref{fig:bragg_results} b), is likely a local minimum. As can be seen on the convergence curves in Fig. \ref{fig:convergencecurve_2} for such a budget, BFGS has converged and since the value of the cost function is higher that what DE can find with a larger budget, this means a local optimum has been found. The performances of the device are interesting, and some regularity can definitely be seen in the structure. We emphasize that a degree of partial regularity is discernible in the outcomes of numerous optimization studies documented in the literature. The reflection peak is larger than a random resonance, which we attribute to the relative regularity of the structure. 

 For a larger budget however, BFGS as well as CMA seem to be stuck in a local minimum, while all the variants of DE get close to the Bragg mirror, the likely optimal solution, which presents the largest reflectance peak of all structures we have tested. We played with variants of BFGS (included in Nevergrad) with restarts: this improves BFGS, but not enough to make it competitive, in particular in Photovoltaics problems.

\begin{figure}
    \centering
    \includegraphics[width=\linewidth]{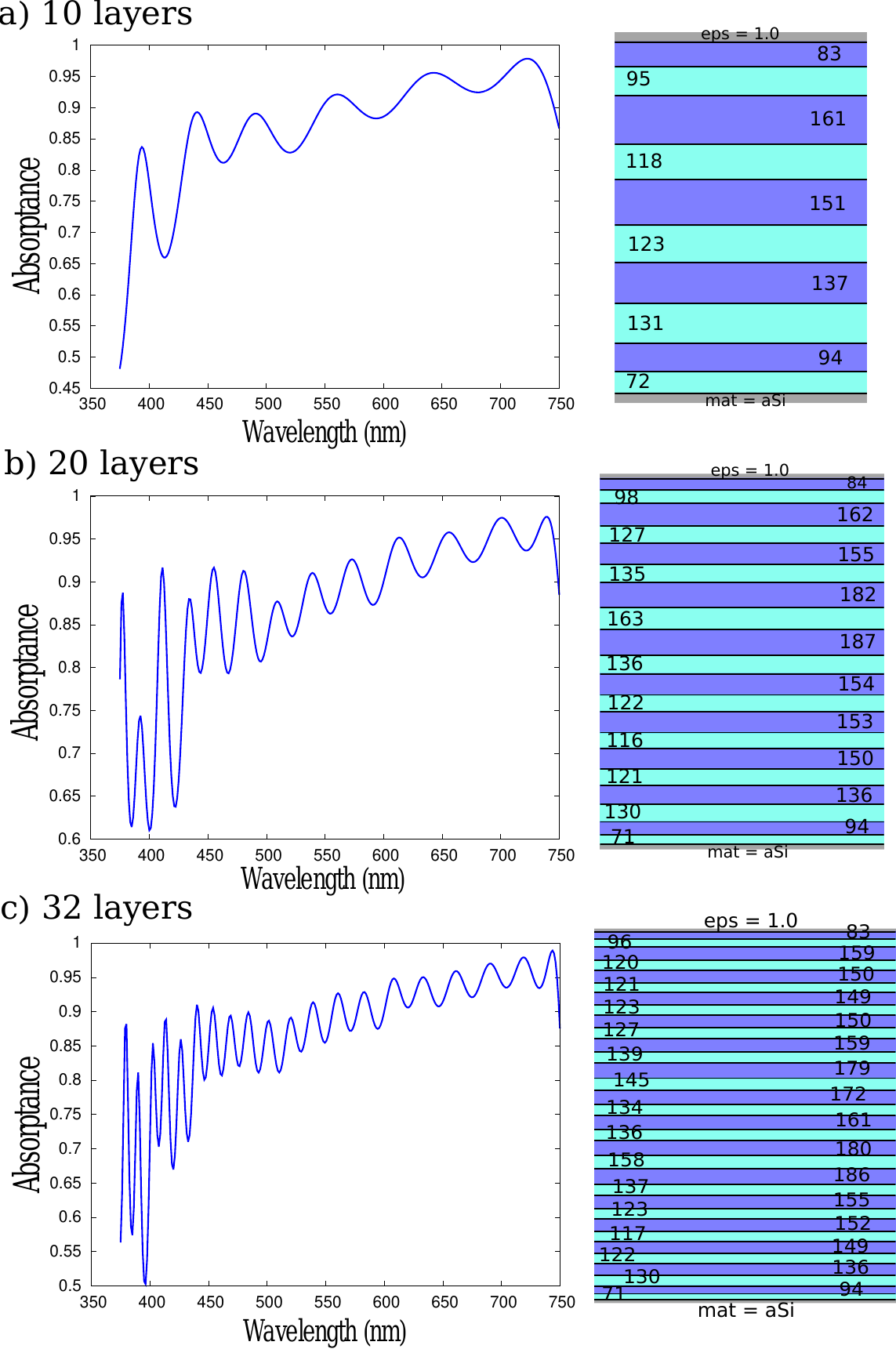}
    \caption{Best structures (right) and their absorptance spectrum (left) obtained for the \texttt{Photolvoltaics} case with {\bf a)} 10 layers   {\bf b)} 20 layers  and {\bf c)} 32 layers. The light, resp. dark color represents the high (3), resp. low (2) permittivity material. The grey color represents substrate and superstrat. 
}
    \label{fig:arpv_results}
\end{figure}

The \texttt{Photovoltaics} case is a perfect example of a modular structure. Figure~\ref{fig:arpv_results} shows the best results for all the \texttt{Photovoltaics} cases, all produced by QODE.
It is important to notice that, whatever the number of layers and even though the structures have been obtained for independent runs, the three upper and up to 5 lower layers are common to all the structures. For the largest number of layers, periodical patterns (alternating 120 nm resp. 150 nm thick layers for permittivity 3.0 resp. 2.0) are appearing. A previous study has shown that the upper and lower layers allow light to be coupled in and out of the central photonic crystal more easily\cite{bennet2021analysis}. The fact that they appear consistently when the number of layers varies indicates their physical importance. They allow to diminish the oscillations characteristics of Bragg mirrors outside the bandgap that can be seen in Fig.~\ref{fig:bragg_results} c) on the reflectance. Such oscillations are detrimental to the absorption.

Regularity emerges in a similar way in the case of the 2D grating we have studied, a problem inspired by the architecture present on the wings of Morpho butterflies\cite{barry2020evolutionary}. As shown in Fig.~\ref{fig:2D_3D_problems}, looking for a way to minimize specular reflection leads to a remarkably regular structure shown in Fig.~\ref{fig:2D_3D_problems}c) even though the structures considered by the algorithms can be different (see Fig. \ref{fig:2D_3D_problems}a). 

Sometimes DE is able to generate structures like chirped dielectric mirrors\cite{barry2020evolutionary}, which locally resemble a Bragg mirror but whose parameters changes gradually and smoothly. Such a structure cannot be considered periodic. It is however modular, as the different parts of the structure are obviously able to reflect different parts of the spectrum. We call this kind of structures regular, even if it is not periodic.
To summarize, a periodic structure is obviously regular, but a regular (\textit{e.g.,} modular) structure is not necessarily periodic.

From the cases presented here and our experience, it can generally be expected that periodic or regular structures will also be the most effective in influencing the behavior of light because it is a wave -- so that it is particularly sensitive to periodicity. The Bragg mirror, for instance, has a higher reflectivity than any disordered structure that we have generated. The thickness values for the AR coating point towards a structure containing a photonic crystal, even if they are not as precise as with the \texttt{Bragg} case, probably because of the irregularities of the solar spectrum. We underline that we have run these optimizations a large number of times on these two cases and never found any better solution that the ones presented here.
This strongly suggests that regularity should be expected and even sought after\cite{gondarenko2006spontaneous} and we underline that in many results from the literature, periodic pattern often seem to emerge spontaneously (see Fig.~\ref{fig:satisfying_structures} below) . 

{\bf Robustness}. In photonics, robustness to fabrication defaults is always desirable. A robust structure presents an efficiency which will not change much when the parameters are slightly modified. From an optimization point of view, this simply means that the attraction basin of the desired solution should be flat. Evaluating the robustness of a solution is thus computationally costly because it involves modifying a large number of parameters and computing the change in the cost function. It can be tempting to include the robustness of the structure in the cost function\cite{robustoptim,robustoptim2}, so that robust solutions appear to have a lower cost function. In our experience, this may lead to an improvement in the quality of the solutions, and produce regular structures more often but at the cost of a much larger computational burden. 

Robustness can be assessed indirectly by examining the spectral response of a structure. This becomes particularly evident in the context of the Bragg mirror. A contraction or dilation of the entire structure primarily shifts the forbidden band and consequently alters the position of the reflectance maximum. As a consequence, there exists a direct correlation between the spectral size of the photonic band-gap and the resilience of the reflectance to structural contractions at the working wavelength. In this context, since the periodicity of a Bragg mirror is what makes its bandgap larger, the connection between regularity and robustness is explicit. While establishing such a link in a universally general manner is impossible, our current findings consistently support this relationship.

\section{Good practices for optimizing photonic structures}
\label{sec:good_practices}

The optimization of photonic structures thus presents quite a few specific characteristics, which influence the strategy to be followed in this particular context. While the computational cost may put a limit to the problems that can be tackled, we give below a list of strategies that, applied together, constitute a methodology for the optimization of photonic structures. 

One of the most important questions is, when we should stop the optimization. It is impossible to prove that a solution is optimal. However, there are ways do determine the quality of a solution. We give below criteria that can help to determine whether a solution is satisfactory -- meaning the optimization can be stopped -- or not.

\subsection{Optimization methodology}

Our methodology consists in maximizing the information extracted from the observables we have defined above, and to make use of specific characteristics of photonic problems to gradually establish confidence in the generated solutions. We leave other strategies that could also make interesting solutions emerge for further work (e.g. multi-objective optimization, or the use of manufacturing noise in the cost function). 

{\bf{Convergence curves for the determination of the budget.}} The presence of plateaus in a convergence curve indicates that the algorithm  has converged, suggesting that the budget is adequate. On the contrary, the absence of plateaus on the consistency curve suggests that the budget should be increased. 

{\bf{Systematic use of consistency curves for checking local minima.}} Even when relying on a single algorithm, since global algorithms are typically non deterministic, it is necessary to launch multiple runs  in order to conduct a statistical analysis with a consistency curve. Ideally, the consistency curve exhibits even a small plateau at its minimum value (see Fig.~\ref{fig:consistencycurve}a). However, when this is not the case (see Fig.~\ref{fig:consistencycurve}f), the solution should be considered with a lot of caution. 

{\bf{Changing the number of parameters.}} In photonics problems, it is generally straightforward to gradually increase the number of elements that can be optimized without changing the nature of the problem. In the cases presented above, this can be done by increasing the number of layers of the structure, but it could for example also be through a decrease in the discretization stepsize, e.g. in topology optimization. Structures with different numbers of layers can then be compared in terms of performances. It can be generally expected that increasing the number of degrees of freedom leads to improved optimized performances. Plotting the minimum value of the cost function as a function of complexity, represented by the number of layers or elements in the structure, can provide valuable insights: e.g. if increasing the number of layers does not improve performance, which indicates that the difficulty of the problem has also increased, continuing that path is likely pointless\cite{bennet2021analysis}.

{\bf{Parametrization bias awareness: meaningful representations make the optimization problem easier.}} 
In parametric optimization, when a handful of parameters are needed to describe a relatively complex device, the choice of these parameters and of their limits (which defines the optimization domain) is crucial. More precisely, these initial choices may introduce biases, favor certain types of algorithms or make the convergence more difficult. When, for instance, the parametrization is chosen such that subsets of parameters correspond to components of the structure, algorithms like DE are particularly efficient. In DE, when a component of a structure is widely spread in the whole population, it might be exactly preserved through the iterations, whereas many other algorithms keep perturbing all variables. 
When the different parameters do not describe a part of the structure but a more global property, other kinds of algorithms might be more relevant, as has been underlined in previous works\cite{schneider2019benchmarking}.

{\bf{Sensitivity to the optimization domain: changing bounds.}} 
In many cases, the imposed constraints strongly control the emerging solutions.
For example, using a medium with an extremely high refractive index (typically infinite) is a simple but not realistic way to reflect light completely. The constraints on the refractive index values are therefore the fundamental reason for the production of Bragg mirrors as a solution. However, there are instances where the constraints become too demanding, making it difficult to find satisfactory solutions. It is important in that case to verify whether some parameters are stuck at the optimization domain boundary (i.e. if the boundary constraints are active). On the other hand, when a satisfactory solution is produced, it can be informative to add or remove constraints or expand the optimization domain. Bragg mirrors tend to emerge, whether the refractive indices are allowed to vary within certain limits or are imposed, with the latter case being straightforward. A clear understanding of the conditions under which a solution is generated also contributes to building confidence.

{\bf{Leverage your physical intuition.}} We underline that in optimization, there are no rules {\em a priori}. If, for instance, it makes sense physically to modify a solution by hand, this should not be considered forbidden or "cheating", especially when it seems that the algorithm is stuck in a local optimum that can be criticized based on a physical reasoning\cite{frellsen_topology_2016}. The limits of the optimization range can also be set to encourage the algorithm to explore areas where promising solutions are likely based on physics, or to stay within specific functional ranges\cite{moreau2012optically}. This approach usually makes the task easier for the algorithm and provides solutions that are easier to understand and thus more satisfactory. Physical intuition is what often determines the conditions in which the optimization takes place and what allows to detect parametrization biases, or even that a problem is not well posed enough for any algorithm to find a satisfactory solution. It should never be overlooked.

\subsection{Assessing the quality of a solution}

Usually, no solution can be proven optimal, due to the impossibility to explore the entire space of possible solutions or to locate all the local minima. Therefore, it becomes necessary to establish criteria that enhance the confidence in a solution.
Besides the optimization criteria above based on optimization observables, a physical analysis is possible, as developped in the present section. When enough confidence in a solution has been built, it can be deemed satisfactory. 

 {\bf Consistency}. A solution that has been obtained at least more than once inspires greater confidence. If it is obtained repeatedly, it might correspond to a plateau of similarly good solutions on the consistency curve of the most efficient algorithm. In that case the solution can be deemed truly consistent and particularly trustworthy. This is most often not the case. When the best solution is obtained only for a single run, this should be considered indicative of a local minimum. We regret that, except in a few cases\cite{su2020nanophotonic}, elements allowing to assess the consistency of a solution, even if this is not a consistency curve, are generally not given.

{\bf Spontaneous emergence of regularity}. In photonics, periodical or regular structures  are ubiquitous. This can be directly attributed to the wave nature of the underlying phenomena. As underlined in a pioneering optimization study\cite{gondarenko2006spontaneous}, ``the emergence of periodicity suggests that periodicity is a principal condition for strong light manipulation.'' Many studies have shown the emergence of partially periodic structures, even when the solution lacks complete periodicity or regularity (like for chirped dielectric mirrors typically\cite{barry2020evolutionary}).
However, when an algorithm proposes completely periodic structures as a solution, they naturally inspire more satisfaction. Based on our experience, we have yet to encounter a simple problem where fully disordered structures outperform regular ones in terms of efficiency.

A symmetrical structure is typically considered as more regular. We believe that the spontaneous emergence of symmetry also reinforces the confidence that can be placed in a solution. We underline that this is rare, as symmetry is often imposed spontaneously in the parametrization -- this tends to simplify problems noticeably.

\begin{figure*}
    \centering
    \includegraphics[width=\linewidth]{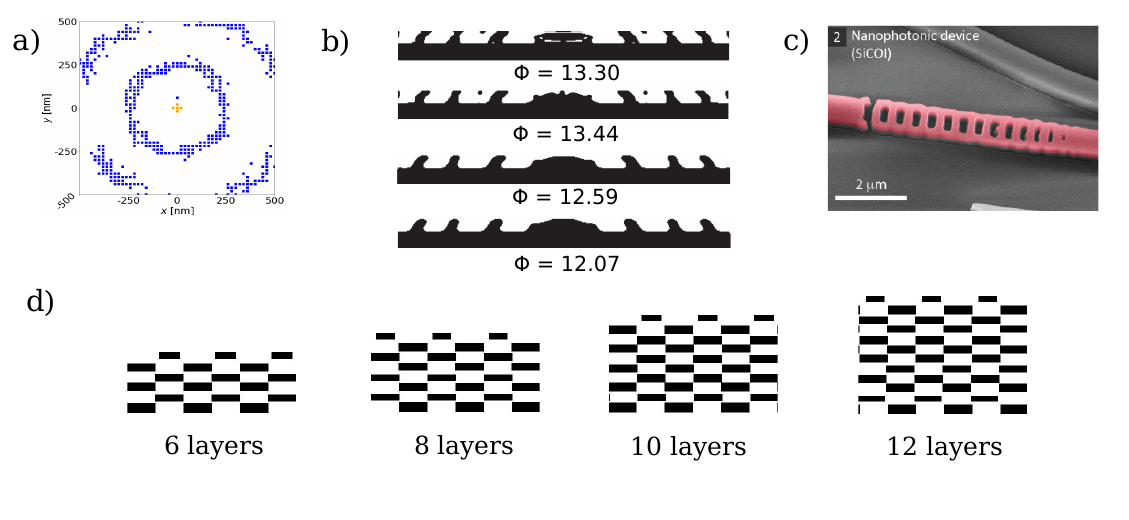}
    \caption{{\bf Spontaneous emergence of regularity or periodicity}. When periodicity or regularity emerge, solutions appear much more satisfactory and are more likely to be analyzed physically.    
    a) When the position of nanopillars of silicon are optimized to enhance the magnetic Purcell factor, using the Differential Evolution algorithm, two ring cavities spontaneously emerge\cite{bruleMagneticElectricPurcell2022}.
    b) Two-dimensional dielectric metalense designs obtained by topology optimization\cite{christiansen2021compact} with a relatively low number of parameters. 
    c) Silicon carbide optical cavity obtained using gradient-based optimization \cite{yang2023inverse} showing a regular pattern (holes of increasing dimensions in the waveguide).
    d) When gratings of rectangular blocks of chitin are optimized to maximize the scattered reflection of a 450~nm wavelength in the first diffraction orders, while minimizing the specular reflection of a range of visible wavelengths quasi perfect checkboards with interdigitated blocks are produced\cite{barry2020evolutionary}.
    }
    \label{fig:satisfying_structures}
\end{figure*}

Overall, we do not necessarily prioritize performance over aesthetics, as both aspects are inherently intertwined in photonics. In the \texttt{Bragg} mirror benchmark, no disordered structure has ever presented a better performance than a Bragg mirror with the same number of layers. In the \texttt{Photovoltaics} case, the irregularities can be linked to the noise in the solar spectrum and, as a consequence, in the cost function. Irregularities in that case improve the performances, but the periodic pattern is still distinguishable and is central for the overall efficiency. In the case of multilayered structures, regularity or periodicity may be more likely to emerge, due to the relative simplicity of the geometry. However, in the literature, relatively regular patterns (in the sense defined above) emerge all the time, as shown in Fig. \ref{fig:satisfying_structures}. Sometimes the patterns look unfinished, perhaps indicating the solution can be further improved -- which is likely if a local algorithm has been used. In our experience, regular or periodic pattern, including spontaneously symmetrical ones, can be generated  also in more complex setups\cite{teytaud2022discrete}.

{\bf Physical interpretability}. The solutions that are most satisfactory  are those that can be readily understood from a physical point of view. We underline that, generally, only periodic, regular or modular structures can be truly understood. This is more difficult for completely disordered structures, except if the disorder itself is tailored, which cannot be ruled out.
The absence of physical interpretability is likely what hinders the widespread adoption of optimization as a daily research tool within the community. Sometimes, the solutions can be studied and fully understood {\em a posteriori}\cite{bennet2021analysis}. Although this does not guarantee optimality, this is at least a good reason to stop looking for alternative solutions: any solution that is comprehensible and understandable can serve as a valuable source of inspiration for manually generated structures and can offer valuable design rules.
In rare situations, algorithms can produce solutions that resemble patterns found in nature, on insect wings for instance, which have evolved for optical purposes. Although these occurrences are uncommon, they can be highly satisfactory, as they align with the concept of evolution as a form of optimization. However, due to their infrequency, they cannot be included in the above criteria. In the case of the reflection problem which we have considered in this work, this criterion is obviously fulfilled too as Bragg mirror are commonly found in nature. 

We underline that these criteria to determine whether a solution is satisfactory or not can be applied to any inverse design problem, whether it is solved by topology optimization, shape optimization, parametric optimization or any other technique. We advise all authors, as far as possible, to publish all the necessary information so that other researchers can reproduce and verify the quality of optimization results. When this is done, particularly interesting and thorough discussions become possible\cite{su2020nanophotonic}.

\section{Conclusion} 

In this paper, we have presented different types of popular optimization algorithms and compared some of them using the Nevergrad platform on three typical photonic problems. We have shown how algorithms can be rigorously and thoroughly compared, relying on specific observables. We proposed a rigorous methodology and offered some advice for conducting high-performance design optimizations and evaluating the quality of a solution. We have illustrated this methodology on a wide range of cases that we think are representative of photonic problems. We have finally provided Jupyter Notebooks that illustrate our workflow and can be used to reproduce the presented benchmarks.

For low-dimensional problems such as ellipsometry or, more generally, those based on the search for a small number of parameters, a lot of algorithms seem adapted, including local optimization algorithms or Bayesian optimization, the latter being particularly useful when evaluating a solution proves costly -- a direction we have not studied here.

For problems that can be considered as inverse design problems, not all approaches are effective. We have shown that photonic problems are characterized by a large number of local minima that render the optimization intrinsically difficult. We have compared our algorithms to what amounts to a steepest descent with random starting points. Except when the problems become too difficult (because of a large number of parameters or because it is noisy), algorithms like Covariance Matrix Adaptation (CMA) or Differential Evolution (DE) constitute a more efficient approach. We tend to recommend DE (and particularly its quasi-oppositional version, QODE) because it proves to be more robust while being simple to implement. DE is never a bad choice, even if it may require a large number of evaluations of the cost function to perform well, and it seems particularly adapted for photonics.

We meant this work as a tutorial, but also as a warning. Optimization is difficult because of its unique curse: it is never possible to guarantee that a solution is optimum, or even close to it, making science much more difficult. This must lead to extra caution, and no solution should ever be deemed optimal, only optimized\cite{christiansen2021compact}.

We underline that the field of optimization is awash in questionable claims of novelty and efficiency. In order to avoid a similar reproducibility crisis in photonics, adopting an open science approach is imperative: data regarding the different runs of optimization should be published, codes should be shared and both should be discussed\cite{su2020nanophotonic,jiang2020metanet}. We are convinced that the optimization of photonics structures is a work intensive domain, and that neither a single method nor a single team will be enough to uncover what optimization can bring in terms of innovation. 

There also is a danger in deeming a solution satisfactory when it should not: to miss innovative and more efficient structures. Given the potential of inverse design but the difficulty to find structures that would be commercially viable\cite{molesky2018inverse}, there is a danger in giving up too soon and miss particularly efficient structures. Fortunately, physical analysis of structures seems to be a powerful tool to discuss both the solutions and the optimization process itself.  

We have shown in the present work how modern numerical tools have made the use of optimization much simpler and efficient in photonics. Even for well defined functioning regimes suggested by physics itself, with a relatively low number of parameters, numerically optimizing a photonic structure often yields unexpectedly high performances. 
We underline that numerical optimization is now able to produce photonics structures that can be understood. This constitutes a complete change compared to the times when inefficient algorithms (such as the first genetic algorithms) were producing disordered and impossible to understand results. Open science approaches now allow any researcher in the field to use such tools easily. We hope that our work will encourage fellow researchers within the community to seamlessly integrate optimization tools into their routine practice and to join the effort in discovering novel and more efficient structures to address the challenges of the future.

\section*{Acknowledgements}

A.M. is an Academy CAP 20-25 chair holder. He acknowledges the support received from the Agence Nationale de la Recherche of the French government through the program Investissements d'Avenir (16-IDEX-0001 CAP 20-25). This work was supported by the International Research Center "Innovation Transportation and Production Systems" of the Clermont-Ferrand I-SITE CAP 20-25.
P.R.W. acknowledges the support of the French Agence Nationale de la Recherche (ANR) under grant ANR-22-CE24-0002 (project NAINOS), and from the Toulouse high performance computing facility CALMIP (grant p20010).

\bibliography{apssamp}

\begin{thebibliography}{92}%
\makeatletter
\providecommand \@ifxundefined [1]{%
 \@ifx{#1\undefined}
}%
\providecommand \@ifnum [1]{%
 \ifnum #1\expandafter \@firstoftwo
 \else \expandafter \@secondoftwo
 \fi
}%
\providecommand \@ifx [1]{%
 \ifx #1\expandafter \@firstoftwo
 \else \expandafter \@secondoftwo
 \fi
}%
\providecommand \natexlab [1]{#1}%
\providecommand \enquote  [1]{``#1''}%
\providecommand \bibnamefont  [1]{#1}%
\providecommand \bibfnamefont [1]{#1}%
\providecommand \citenamefont [1]{#1}%
\providecommand \href@noop [0]{\@secondoftwo}%
\providecommand \href [0]{\begingroup \@sanitize@url \@href}%
\providecommand \@href[1]{\@@startlink{#1}\@@href}%
\providecommand \@@href[1]{\endgroup#1\@@endlink}%
\providecommand \@sanitize@url [0]{\catcode `\\12\catcode `\$12\catcode `\&12\catcode `\#12\catcode `\^12\catcode `\_12\catcode `\%12\relax}%
\providecommand \@@startlink[1]{}%
\providecommand \@@endlink[0]{}%
\providecommand \url  [0]{\begingroup\@sanitize@url \@url }%
\providecommand \@url [1]{\endgroup\@href {#1}{\urlprefix }}%
\providecommand \urlprefix  [0]{URL }%
\providecommand \Eprint [0]{\href }%
\providecommand \doibase [0]{https://doi.org/}%
\providecommand \selectlanguage [0]{\@gobble}%
\providecommand \bibinfo  [0]{\@secondoftwo}%
\providecommand \bibfield  [0]{\@secondoftwo}%
\providecommand \translation [1]{[#1]}%
\providecommand \BibitemOpen [0]{}%
\providecommand \bibitemStop [0]{}%
\providecommand \bibitemNoStop [0]{.\EOS\space}%
\providecommand \EOS [0]{\spacefactor3000\relax}%
\providecommand \BibitemShut  [1]{\csname bibitem#1\endcsname}%
\let\auto@bib@innerbib\@empty
\bibitem [{\citenamefont {Molesky}\ \emph {et~al.}(2018)\citenamefont {Molesky}, \citenamefont {Lin}, \citenamefont {Piggott}, \citenamefont {Jin}, \citenamefont {Vuckovi{\'c}},\ and\ \citenamefont {Rodriguez}}]{molesky2018inverse}%
  \BibitemOpen
  \bibfield  {author} {\bibinfo {author} {\bibfnamefont {S.}~\bibnamefont {Molesky}}, \bibinfo {author} {\bibfnamefont {Z.}~\bibnamefont {Lin}}, \bibinfo {author} {\bibfnamefont {A.~Y.}\ \bibnamefont {Piggott}}, \bibinfo {author} {\bibfnamefont {W.}~\bibnamefont {Jin}}, \bibinfo {author} {\bibfnamefont {J.}~\bibnamefont {Vuckovi{\'c}}},\ and\ \bibinfo {author} {\bibfnamefont {A.~W.}\ \bibnamefont {Rodriguez}},\ }\href@noop {} {\bibfield  {journal} {\bibinfo  {journal} {Nature Photonics}\ }\textbf {\bibinfo {volume} {12}},\ \bibinfo {pages} {659} (\bibinfo {year} {2018})}\BibitemShut {NoStop}%
\bibitem [{\citenamefont {Campbell}\ \emph {et~al.}(2019)\citenamefont {Campbell}, \citenamefont {Sell}, \citenamefont {Jenkins}, \citenamefont {Whiting}, \citenamefont {Fan},\ and\ \citenamefont {Werner}}]{campbell2019review}%
  \BibitemOpen
  \bibfield  {author} {\bibinfo {author} {\bibfnamefont {S.~D.}\ \bibnamefont {Campbell}}, \bibinfo {author} {\bibfnamefont {D.}~\bibnamefont {Sell}}, \bibinfo {author} {\bibfnamefont {R.~P.}\ \bibnamefont {Jenkins}}, \bibinfo {author} {\bibfnamefont {E.~B.}\ \bibnamefont {Whiting}}, \bibinfo {author} {\bibfnamefont {J.~A.}\ \bibnamefont {Fan}},\ and\ \bibinfo {author} {\bibfnamefont {D.~H.}\ \bibnamefont {Werner}},\ }\href@noop {} {\bibfield  {journal} {\bibinfo  {journal} {Optical Materials Express}\ }\textbf {\bibinfo {volume} {9}},\ \bibinfo {pages} {1842} (\bibinfo {year} {2019})}\BibitemShut {NoStop}%
\bibitem [{\citenamefont {Elsawy}\ \emph {et~al.}(2020)\citenamefont {Elsawy}, \citenamefont {Lanteri}, \citenamefont {Duvigneau}, \citenamefont {Fan},\ and\ \citenamefont {Genevet}}]{elsawy2020numerical}%
  \BibitemOpen
  \bibfield  {author} {\bibinfo {author} {\bibfnamefont {M.~M.}\ \bibnamefont {Elsawy}}, \bibinfo {author} {\bibfnamefont {S.}~\bibnamefont {Lanteri}}, \bibinfo {author} {\bibfnamefont {R.}~\bibnamefont {Duvigneau}}, \bibinfo {author} {\bibfnamefont {J.~A.}\ \bibnamefont {Fan}},\ and\ \bibinfo {author} {\bibfnamefont {P.}~\bibnamefont {Genevet}},\ }\href@noop {} {\bibfield  {journal} {\bibinfo  {journal} {Laser \& Photonics Reviews}\ }\textbf {\bibinfo {volume} {14}},\ \bibinfo {pages} {1900445} (\bibinfo {year} {2020})}\BibitemShut {NoStop}%
\bibitem [{\citenamefont {Chen}\ \emph {et~al.}(2022)\citenamefont {Chen}, \citenamefont {Jiang},\ and\ \citenamefont {Fan}}]{chen2022algorithm}%
  \BibitemOpen
  \bibfield  {author} {\bibinfo {author} {\bibfnamefont {M.}~\bibnamefont {Chen}}, \bibinfo {author} {\bibfnamefont {J.}~\bibnamefont {Jiang}},\ and\ \bibinfo {author} {\bibfnamefont {J.~A.}\ \bibnamefont {Fan}},\ }\href@noop {} {\bibfield  {journal} {\bibinfo  {journal} {ACS Photonics}\ }\textbf {\bibinfo {volume} {9}},\ \bibinfo {pages} {2860} (\bibinfo {year} {2022})}\BibitemShut {NoStop}%
\bibitem [{\citenamefont {Bendsoe}\ and\ \citenamefont {Sigmund}(2003)}]{bendsoe2003topology}%
  \BibitemOpen
  \bibfield  {author} {\bibinfo {author} {\bibfnamefont {M.~P.}\ \bibnamefont {Bendsoe}}\ and\ \bibinfo {author} {\bibfnamefont {O.}~\bibnamefont {Sigmund}},\ }\href@noop {} {\emph {\bibinfo {title} {Topology optimization: theory, methods, and applications}}}\ (\bibinfo  {publisher} {Springer Science \& Business Media},\ \bibinfo {year} {2003})\BibitemShut {NoStop}%
\bibitem [{\citenamefont {Sigmund}(2011)}]{sigmund2011usefulness}%
  \BibitemOpen
  \bibfield  {author} {\bibinfo {author} {\bibfnamefont {O.}~\bibnamefont {Sigmund}},\ }\href@noop {} {\bibfield  {journal} {\bibinfo  {journal} {Structural and Multidisciplinary Optimization}\ }\textbf {\bibinfo {volume} {43}},\ \bibinfo {pages} {589} (\bibinfo {year} {2011})}\BibitemShut {NoStop}%
\bibitem [{\citenamefont {Barry}\ \emph {et~al.}(2020)\citenamefont {Barry}, \citenamefont {Berthier}, \citenamefont {Wilts}, \citenamefont {Cambourieux}, \citenamefont {Bennet}, \citenamefont {Poll{\`e}s}, \citenamefont {Teytaud}, \citenamefont {Centeno}, \citenamefont {Biais},\ and\ \citenamefont {Moreau}}]{barry2020evolutionary}%
  \BibitemOpen
  \bibfield  {author} {\bibinfo {author} {\bibfnamefont {M.~A.}\ \bibnamefont {Barry}}, \bibinfo {author} {\bibfnamefont {V.}~\bibnamefont {Berthier}}, \bibinfo {author} {\bibfnamefont {B.~D.}\ \bibnamefont {Wilts}}, \bibinfo {author} {\bibfnamefont {M.-C.}\ \bibnamefont {Cambourieux}}, \bibinfo {author} {\bibfnamefont {P.}~\bibnamefont {Bennet}}, \bibinfo {author} {\bibfnamefont {R.}~\bibnamefont {Poll{\`e}s}}, \bibinfo {author} {\bibfnamefont {O.}~\bibnamefont {Teytaud}}, \bibinfo {author} {\bibfnamefont {E.}~\bibnamefont {Centeno}}, \bibinfo {author} {\bibfnamefont {N.}~\bibnamefont {Biais}},\ and\ \bibinfo {author} {\bibfnamefont {A.}~\bibnamefont {Moreau}},\ }\href {https://doi.org/10.1038/s41598-020-68719-3} {\bibfield  {journal} {\bibinfo  {journal} {Scientific reports}\ }\textbf {\bibinfo {volume} {10}},\ \bibinfo {pages} {1} (\bibinfo {year} {2020})}\BibitemShut {NoStop}%
\bibitem [{\citenamefont {Bennet}\ \emph {et~al.}(2021)\citenamefont {Bennet}, \citenamefont {Juillet}, \citenamefont {Ibrahim}, \citenamefont {Berthier}, \citenamefont {Barry}, \citenamefont {R{\'e}veret}, \citenamefont {Bousquet}, \citenamefont {Teytaud}, \citenamefont {Centeno},\ and\ \citenamefont {Moreau}}]{bennet2021analysis}%
  \BibitemOpen
  \bibfield  {author} {\bibinfo {author} {\bibfnamefont {P.}~\bibnamefont {Bennet}}, \bibinfo {author} {\bibfnamefont {P.}~\bibnamefont {Juillet}}, \bibinfo {author} {\bibfnamefont {S.}~\bibnamefont {Ibrahim}}, \bibinfo {author} {\bibfnamefont {V.}~\bibnamefont {Berthier}}, \bibinfo {author} {\bibfnamefont {M.~A.}\ \bibnamefont {Barry}}, \bibinfo {author} {\bibfnamefont {F.}~\bibnamefont {R{\'e}veret}}, \bibinfo {author} {\bibfnamefont {A.}~\bibnamefont {Bousquet}}, \bibinfo {author} {\bibfnamefont {O.}~\bibnamefont {Teytaud}}, \bibinfo {author} {\bibfnamefont {E.}~\bibnamefont {Centeno}},\ and\ \bibinfo {author} {\bibfnamefont {A.}~\bibnamefont {Moreau}},\ }\href@noop {} {\bibfield  {journal} {\bibinfo  {journal} {Physical Review B}\ }\textbf {\bibinfo {volume} {103}},\ \bibinfo {pages} {125135} (\bibinfo {year} {2021})}\BibitemShut {NoStop}%
\bibitem [{\citenamefont {Sörensen}(2015)}]{sorensenmetaheuristicsmetaphorexposed2015}%
  \BibitemOpen
  \bibfield  {author} {\bibinfo {author} {\bibfnamefont {K.}~\bibnamefont {Sörensen}},\ }\href {https://doi.org/https://doi.org/10.1111/itor.12001} {\bibfield  {journal} {\bibinfo  {journal} {International Transactions in Operational Research}\ }\textbf {\bibinfo {volume} {22}},\ \bibinfo {pages} {3} (\bibinfo {year} {2015})},\ \Eprint {https://arxiv.org/abs/https://onlinelibrary.wiley.com/doi/pdf/10.1111/itor.12001} {https://onlinelibrary.wiley.com/doi/pdf/10.1111/itor.12001} \BibitemShut {NoStop}%
\bibitem [{pym(2023)}]{pymoosh}%
  \BibitemOpen
  \href@noop {} {\bibinfo {title} {Pymoosh}},\ \bibinfo {howpublished} {\url{https://pypi.org/project/PyMoosh/}} (\bibinfo {year} {2023}),\ \bibinfo {note} {verified: 2022-06-10}\BibitemShut {NoStop}%
\bibitem [{\citenamefont {Rapin}\ and\ \citenamefont {Teytaud}(2018)}]{nevergrad}%
  \BibitemOpen
  \bibfield  {author} {\bibinfo {author} {\bibfnamefont {J.}~\bibnamefont {Rapin}}\ and\ \bibinfo {author} {\bibfnamefont {O.}~\bibnamefont {Teytaud}},\ }\href@noop {} {\bibinfo {title} {{Nevergrad - A gradient-free optimization platform}}},\ \bibinfo {howpublished} {\url{https://GitHub.com/FacebookResearch/Nevergrad}} (\bibinfo {year} {2018})\BibitemShut {NoStop}%
\bibitem [{\citenamefont {Bennet}\ \emph {et~al.}(2023)\citenamefont {Bennet}, \citenamefont {Teytaud}, \citenamefont {Wiecha}, \citenamefont {Langevin}, \citenamefont {Moreau},\ and\ \citenamefont {Khaireh-Walieh}}]{bennet_2023_10228377}%
  \BibitemOpen
  \bibfield  {author} {\bibinfo {author} {\bibfnamefont {P.}~\bibnamefont {Bennet}}, \bibinfo {author} {\bibfnamefont {O.}~\bibnamefont {Teytaud}}, \bibinfo {author} {\bibfnamefont {P.}~\bibnamefont {Wiecha}}, \bibinfo {author} {\bibfnamefont {D.}~\bibnamefont {Langevin}}, \bibinfo {author} {\bibfnamefont {A.}~\bibnamefont {Moreau}},\ and\ \bibinfo {author} {\bibfnamefont {A.}~\bibnamefont {Khaireh-Walieh}},\ }\href {https://doi.org/10.5281/zenodo.10246032} {\bibinfo {title} {Ellawin/tuto\_global\_optimization\_photonics: v2}},\ \bibinfo {howpublished} {\url{https://doi.org/10.5281/zenodo.10246032}} (\bibinfo {year} {2023})\BibitemShut {NoStop}%
\bibitem [{\citenamefont {Wang}\ \emph {et~al.}(2002)\citenamefont {Wang}, \citenamefont {Lu},\ and\ \citenamefont {He}}]{wang2002optimal}%
  \BibitemOpen
  \bibfield  {author} {\bibinfo {author} {\bibfnamefont {Q.}~\bibnamefont {Wang}}, \bibinfo {author} {\bibfnamefont {J.}~\bibnamefont {Lu}},\ and\ \bibinfo {author} {\bibfnamefont {S.}~\bibnamefont {He}},\ }\href@noop {} {\bibfield  {journal} {\bibinfo  {journal} {Applied optics}\ }\textbf {\bibinfo {volume} {41}},\ \bibinfo {pages} {7644} (\bibinfo {year} {2002})}\BibitemShut {NoStop}%
\bibitem [{\citenamefont {Gondarenko}\ \emph {et~al.}(2006)\citenamefont {Gondarenko}, \citenamefont {Preble}, \citenamefont {Robinson}, \citenamefont {Chen}, \citenamefont {Lipson},\ and\ \citenamefont {Lipson}}]{gondarenko2006spontaneous}%
  \BibitemOpen
  \bibfield  {author} {\bibinfo {author} {\bibfnamefont {A.}~\bibnamefont {Gondarenko}}, \bibinfo {author} {\bibfnamefont {S.}~\bibnamefont {Preble}}, \bibinfo {author} {\bibfnamefont {J.}~\bibnamefont {Robinson}}, \bibinfo {author} {\bibfnamefont {L.}~\bibnamefont {Chen}}, \bibinfo {author} {\bibfnamefont {H.}~\bibnamefont {Lipson}},\ and\ \bibinfo {author} {\bibfnamefont {M.}~\bibnamefont {Lipson}},\ }\href@noop {} {\bibfield  {journal} {\bibinfo  {journal} {Physical review letters}\ }\textbf {\bibinfo {volume} {96}},\ \bibinfo {pages} {143904} (\bibinfo {year} {2006})}\BibitemShut {NoStop}%
\bibitem [{\citenamefont {Piggott}\ \emph {et~al.}(2015)\citenamefont {Piggott}, \citenamefont {Lu}, \citenamefont {Lagoudakis}, \citenamefont {Petykiewicz}, \citenamefont {Babinec},\ and\ \citenamefont {Vu{\v{c}}kovi{\'c}}}]{piggott2015inverse}%
  \BibitemOpen
  \bibfield  {author} {\bibinfo {author} {\bibfnamefont {A.~Y.}\ \bibnamefont {Piggott}}, \bibinfo {author} {\bibfnamefont {J.}~\bibnamefont {Lu}}, \bibinfo {author} {\bibfnamefont {K.~G.}\ \bibnamefont {Lagoudakis}}, \bibinfo {author} {\bibfnamefont {J.}~\bibnamefont {Petykiewicz}}, \bibinfo {author} {\bibfnamefont {T.~M.}\ \bibnamefont {Babinec}},\ and\ \bibinfo {author} {\bibfnamefont {J.}~\bibnamefont {Vu{\v{c}}kovi{\'c}}},\ }\href@noop {} {\bibfield  {journal} {\bibinfo  {journal} {Nature Photonics}\ }\textbf {\bibinfo {volume} {9}},\ \bibinfo {pages} {374} (\bibinfo {year} {2015})}\BibitemShut {NoStop}%
\bibitem [{\citenamefont {Einarsson}\ \emph {et~al.}(2019)\citenamefont {Einarsson}, \citenamefont {Gauy}, \citenamefont {Lengler}, \citenamefont {Meier}, \citenamefont {Mujika}, \citenamefont {Steger},\ and\ \citenamefont {Weissenberger}}]{relengler}%
  \BibitemOpen
  \bibfield  {author} {\bibinfo {author} {\bibfnamefont {H.}~\bibnamefont {Einarsson}}, \bibinfo {author} {\bibfnamefont {M.~M.}\ \bibnamefont {Gauy}}, \bibinfo {author} {\bibfnamefont {J.}~\bibnamefont {Lengler}}, \bibinfo {author} {\bibfnamefont {F.}~\bibnamefont {Meier}}, \bibinfo {author} {\bibfnamefont {A.}~\bibnamefont {Mujika}}, \bibinfo {author} {\bibfnamefont {A.}~\bibnamefont {Steger}},\ and\ \bibinfo {author} {\bibfnamefont {F.}~\bibnamefont {Weissenberger}},\ }\href@noop {} {\bibfield  {journal} {\bibinfo  {journal} {Theor. Comput. Sci.}\ }\textbf {\bibinfo {volume} {785}},\ \bibinfo {pages} {150} (\bibinfo {year} {2019})}\BibitemShut {NoStop}%
\bibitem [{\citenamefont {Doerr}\ and\ \citenamefont {Neumann}(2021)}]{doerr2021survey}%
  \BibitemOpen
  \bibfield  {author} {\bibinfo {author} {\bibfnamefont {B.}~\bibnamefont {Doerr}}\ and\ \bibinfo {author} {\bibfnamefont {F.}~\bibnamefont {Neumann}},\ }\href@noop {} {\bibfield  {journal} {\bibinfo  {journal} {ACM Transactions on Evolutionary Learning and Optimization}\ }\textbf {\bibinfo {volume} {1}},\ \bibinfo {pages} {1} (\bibinfo {year} {2021})}\BibitemShut {NoStop}%
\bibitem [{\citenamefont {B{\"a}ck}\ \emph {et~al.}(2023)\citenamefont {B{\"a}ck}, \citenamefont {Kononova}, \citenamefont {van Stein}, \citenamefont {Wang}, \citenamefont {Antonov}, \citenamefont {Kalkreuth}, \citenamefont {de~Nobel}, \citenamefont {Vermetten}, \citenamefont {de~Winter},\ and\ \citenamefont {Ye}}]{back2023evolutionary}%
  \BibitemOpen
  \bibfield  {author} {\bibinfo {author} {\bibfnamefont {T.~H.}\ \bibnamefont {B{\"a}ck}}, \bibinfo {author} {\bibfnamefont {A.~V.}\ \bibnamefont {Kononova}}, \bibinfo {author} {\bibfnamefont {B.}~\bibnamefont {van Stein}}, \bibinfo {author} {\bibfnamefont {H.}~\bibnamefont {Wang}}, \bibinfo {author} {\bibfnamefont {K.~A.}\ \bibnamefont {Antonov}}, \bibinfo {author} {\bibfnamefont {R.~T.}\ \bibnamefont {Kalkreuth}}, \bibinfo {author} {\bibfnamefont {J.}~\bibnamefont {de~Nobel}}, \bibinfo {author} {\bibfnamefont {D.}~\bibnamefont {Vermetten}}, \bibinfo {author} {\bibfnamefont {R.}~\bibnamefont {de~Winter}},\ and\ \bibinfo {author} {\bibfnamefont {F.}~\bibnamefont {Ye}},\ }\href@noop {} {\bibfield  {journal} {\bibinfo  {journal} {Evolutionary Computation}\ }\textbf {\bibinfo {volume} {31}},\ \bibinfo {pages} {81} (\bibinfo {year} {2023})}\BibitemShut {NoStop}%
\bibitem [{\citenamefont {Gagnon}\ \emph {et~al.}(2013)\citenamefont {Gagnon}, \citenamefont {Dumont},\ and\ \citenamefont {Dub{\'e}}}]{gagnon2013multiobjective}%
  \BibitemOpen
  \bibfield  {author} {\bibinfo {author} {\bibfnamefont {D.}~\bibnamefont {Gagnon}}, \bibinfo {author} {\bibfnamefont {J.}~\bibnamefont {Dumont}},\ and\ \bibinfo {author} {\bibfnamefont {L.~J.}\ \bibnamefont {Dub{\'e}}},\ }\href@noop {} {\bibfield  {journal} {\bibinfo  {journal} {Optics Letters}\ }\textbf {\bibinfo {volume} {38}},\ \bibinfo {pages} {2181} (\bibinfo {year} {2013})}\BibitemShut {NoStop}%
\bibitem [{\citenamefont {Teytaud}\ \emph {et~al.}(2022)\citenamefont {Teytaud}, \citenamefont {Bennet},\ and\ \citenamefont {Moreau}}]{teytaud2022discrete}%
  \BibitemOpen
  \bibfield  {author} {\bibinfo {author} {\bibfnamefont {O.}~\bibnamefont {Teytaud}}, \bibinfo {author} {\bibfnamefont {P.}~\bibnamefont {Bennet}},\ and\ \bibinfo {author} {\bibfnamefont {A.}~\bibnamefont {Moreau}},\ }\href {https://doi.org/https://doi.org/10.1016/j.photonics.2022.101072} {\bibfield  {journal} {\bibinfo  {journal} {Photonics and Nanostructures - Fundamentals and Applications}\ }\textbf {\bibinfo {volume} {52}},\ \bibinfo {pages} {101072} (\bibinfo {year} {2022})}\BibitemShut {NoStop}%
\bibitem [{\citenamefont {Wu}\ \emph {et~al.}(2017)\citenamefont {Wu}, \citenamefont {Poloczek}, \citenamefont {Wilson},\ and\ \citenamefont {Frazier}}]{wu2017bayesian}%
  \BibitemOpen
  \bibfield  {author} {\bibinfo {author} {\bibfnamefont {J.}~\bibnamefont {Wu}}, \bibinfo {author} {\bibfnamefont {M.}~\bibnamefont {Poloczek}}, \bibinfo {author} {\bibfnamefont {A.~G.}\ \bibnamefont {Wilson}},\ and\ \bibinfo {author} {\bibfnamefont {P.}~\bibnamefont {Frazier}},\ }\href@noop {} {\bibfield  {journal} {\bibinfo  {journal} {Advances in neural information processing systems}\ }\textbf {\bibinfo {volume} {30}} (\bibinfo {year} {2017})}\BibitemShut {NoStop}%
\bibitem [{\citenamefont {Garcia-Santiago}\ \emph {et~al.}(2021)\citenamefont {Garcia-Santiago}, \citenamefont {Burger}, \citenamefont {Rockstuhl},\ and\ \citenamefont {Schneider}}]{garcia2021bayesian}%
  \BibitemOpen
  \bibfield  {author} {\bibinfo {author} {\bibfnamefont {X.}~\bibnamefont {Garcia-Santiago}}, \bibinfo {author} {\bibfnamefont {S.}~\bibnamefont {Burger}}, \bibinfo {author} {\bibfnamefont {C.}~\bibnamefont {Rockstuhl}},\ and\ \bibinfo {author} {\bibfnamefont {P.-I.}\ \bibnamefont {Schneider}},\ }\href@noop {} {\bibfield  {journal} {\bibinfo  {journal} {Journal of Lightwave Technology}\ }\textbf {\bibinfo {volume} {39}},\ \bibinfo {pages} {167} (\bibinfo {year} {2021})}\BibitemShut {NoStop}%
\bibitem [{\citenamefont {Jakšić}\ \emph {et~al.}(2023)\citenamefont {Jakšić}, \citenamefont {Devi}, \citenamefont {Jakšić},\ and\ \citenamefont {Guha}}]{biomimetics8030278}%
  \BibitemOpen
  \bibfield  {author} {\bibinfo {author} {\bibfnamefont {Z.}~\bibnamefont {Jakšić}}, \bibinfo {author} {\bibfnamefont {S.}~\bibnamefont {Devi}}, \bibinfo {author} {\bibfnamefont {O.}~\bibnamefont {Jakšić}},\ and\ \bibinfo {author} {\bibfnamefont {K.}~\bibnamefont {Guha}},\ }\bibfield  {journal} {\bibinfo  {journal} {Biomimetics}\ }\textbf {\bibinfo {volume} {8}},\ \href {https://doi.org/10.3390/biomimetics8030278} {10.3390/biomimetics8030278} (\bibinfo {year} {2023})\BibitemShut {NoStop}%
\bibitem [{\citenamefont {Wolpert}\ and\ \citenamefont {Macready}(1997)}]{wolpert1997nofree}%
  \BibitemOpen
  \bibfield  {author} {\bibinfo {author} {\bibfnamefont {D.}~\bibnamefont {Wolpert}}\ and\ \bibinfo {author} {\bibfnamefont {W.}~\bibnamefont {Macready}},\ }\href {https://doi.org/10.1109/4235.585893} {\bibfield  {journal} {\bibinfo  {journal} {IEEE Transactions on Evolutionary Computation}\ }\textbf {\bibinfo {volume} {1}},\ \bibinfo {pages} {67} (\bibinfo {year} {1997})}\BibitemShut {NoStop}%
\bibitem [{\citenamefont {Markov}(2023)}]{rlgoogle}%
  \BibitemOpen
  \bibfield  {author} {\bibinfo {author} {\bibfnamefont {I.~L.}\ \bibnamefont {Markov}},\ }\href@noop {} {\bibinfo {title} {The false dawn: Reevaluating google's reinforcement learning for chip macro placement}} (\bibinfo {year} {2023}),\ \Eprint {https://arxiv.org/abs/2306.09633} {arXiv:2306.09633 [cs.LG]} \BibitemShut {NoStop}%
\bibitem [{\citenamefont {Gould}\ \emph {et~al.}(2003)\citenamefont {Gould}, \citenamefont {Orban},\ and\ \citenamefont {Toint}}]{gould2003cuter}%
  \BibitemOpen
  \bibfield  {author} {\bibinfo {author} {\bibfnamefont {N.~I.}\ \bibnamefont {Gould}}, \bibinfo {author} {\bibfnamefont {D.}~\bibnamefont {Orban}},\ and\ \bibinfo {author} {\bibfnamefont {P.~L.}\ \bibnamefont {Toint}},\ }\href@noop {} {\bibfield  {journal} {\bibinfo  {journal} {ACM Transactions on Mathematical Software (TOMS)}\ }\textbf {\bibinfo {volume} {29}},\ \bibinfo {pages} {373} (\bibinfo {year} {2003})}\BibitemShut {NoStop}%
\bibitem [{\citenamefont {Gould}\ \emph {et~al.}(2015)\citenamefont {Gould}, \citenamefont {Orban},\ and\ \citenamefont {Toint}}]{gould2015cutest}%
  \BibitemOpen
  \bibfield  {author} {\bibinfo {author} {\bibfnamefont {N.~I.}\ \bibnamefont {Gould}}, \bibinfo {author} {\bibfnamefont {D.}~\bibnamefont {Orban}},\ and\ \bibinfo {author} {\bibfnamefont {P.~L.}\ \bibnamefont {Toint}},\ }\href@noop {} {\bibfield  {journal} {\bibinfo  {journal} {Computational optimization and applications}\ }\textbf {\bibinfo {volume} {60}},\ \bibinfo {pages} {545} (\bibinfo {year} {2015})}\BibitemShut {NoStop}%
\bibitem [{\citenamefont {{DIMACS}}(2021)}]{dimacs}%
  \BibitemOpen
  \bibfield  {author} {\bibinfo {author} {\bibnamefont {{DIMACS}}},\ }\href@noop {} {\bibinfo {title} {Dimacs implementation challenge}},\ \bibinfo {howpublished} {{http://dimacs.rutgers.edu/programs/challenge}} (\bibinfo {year} {2021})\BibitemShut {NoStop}%
\bibitem [{\citenamefont {Hansen}\ \emph {et~al.}(2009)\citenamefont {Hansen}, \citenamefont {Auger}, \citenamefont {Finck},\ and\ \citenamefont {Ros}}]{bbob}%
  \BibitemOpen
  \bibfield  {author} {\bibinfo {author} {\bibfnamefont {N.}~\bibnamefont {Hansen}}, \bibinfo {author} {\bibfnamefont {A.}~\bibnamefont {Auger}}, \bibinfo {author} {\bibfnamefont {S.}~\bibnamefont {Finck}},\ and\ \bibinfo {author} {\bibfnamefont {R.}~\bibnamefont {Ros}},\ }\href@noop {} {\emph {\bibinfo {title} {Real-Parameter Black-Box Optimization Benchmarking 2009: Experimental setup}}},\ \bibinfo {type} {Tech. Rep.}\ \bibinfo {number} {RR-6828}\ (\bibinfo  {institution} {INRIA, France},\ \bibinfo {year} {2009})\BibitemShut {NoStop}%
\bibitem [{\citenamefont {Li}\ \emph {et~al.}(2013)\citenamefont {Li}, \citenamefont {Tang}, \citenamefont {Omidvar}, \citenamefont {Yang},\ and\ \citenamefont {Qin}}]{lsgo}%
  \BibitemOpen
  \bibfield  {author} {\bibinfo {author} {\bibfnamefont {X.}~\bibnamefont {Li}}, \bibinfo {author} {\bibfnamefont {K.}~\bibnamefont {Tang}}, \bibinfo {author} {\bibfnamefont {M.~N.}\ \bibnamefont {Omidvar}}, \bibinfo {author} {\bibfnamefont {Z.}~\bibnamefont {Yang}},\ and\ \bibinfo {author} {\bibfnamefont {K.}~\bibnamefont {Qin}},\ }in\ \href@noop {} {\emph {\bibinfo {booktitle} {CEC 2013 proceedings}}}\ (\bibinfo {year} {2013})\BibitemShut {NoStop}%
\bibitem [{\citenamefont {Gallagher}\ and\ \citenamefont {Saleem}(2018)}]{mlda}%
  \BibitemOpen
  \bibfield  {author} {\bibinfo {author} {\bibfnamefont {M.}~\bibnamefont {Gallagher}}\ and\ \bibinfo {author} {\bibfnamefont {S.}~\bibnamefont {Saleem}},\ }in\ \href@noop {} {\emph {\bibinfo {booktitle} {PPSN'18 workshop}}}\ (\bibinfo {year} {2018})\BibitemShut {NoStop}%
\bibitem [{\citenamefont {Häse}\ \emph {et~al.}(2021)\citenamefont {Häse}, \citenamefont {Aldeghi}, \citenamefont {Hickman}, \citenamefont {Roch}, \citenamefont {Christensen}, \citenamefont {Liles}, \citenamefont {Hein},\ and\ \citenamefont {Aspuru-Guzik}}]{olympus}%
  \BibitemOpen
  \bibfield  {author} {\bibinfo {author} {\bibfnamefont {F.}~\bibnamefont {Häse}}, \bibinfo {author} {\bibfnamefont {M.}~\bibnamefont {Aldeghi}}, \bibinfo {author} {\bibfnamefont {R.~J.}\ \bibnamefont {Hickman}}, \bibinfo {author} {\bibfnamefont {L.~M.}\ \bibnamefont {Roch}}, \bibinfo {author} {\bibfnamefont {M.}~\bibnamefont {Christensen}}, \bibinfo {author} {\bibfnamefont {E.}~\bibnamefont {Liles}}, \bibinfo {author} {\bibfnamefont {J.~E.}\ \bibnamefont {Hein}},\ and\ \bibinfo {author} {\bibfnamefont {A.}~\bibnamefont {Aspuru-Guzik}},\ }\href@noop {} {\bibfield  {journal} {\bibinfo  {journal} {Machine Learning: Science and Technology}\ }\textbf {\bibinfo {volume} {2}},\ \bibinfo {pages} {035021} (\bibinfo {year} {2021})}\BibitemShut {NoStop}%
\bibitem [{\citenamefont {Lee}\ \emph {et~al.}(2018)\citenamefont {Lee}, \citenamefont {Grey}, \citenamefont {Ha}, \citenamefont {Kunz}, \citenamefont {Jain}, \citenamefont {Ye}, \citenamefont {Srinivasa}, \citenamefont {Stilman},\ and\ \citenamefont {Liu}}]{dart}%
  \BibitemOpen
  \bibfield  {author} {\bibinfo {author} {\bibfnamefont {J.}~\bibnamefont {Lee}}, \bibinfo {author} {\bibfnamefont {M.~X.}\ \bibnamefont {Grey}}, \bibinfo {author} {\bibfnamefont {S.}~\bibnamefont {Ha}}, \bibinfo {author} {\bibfnamefont {T.}~\bibnamefont {Kunz}}, \bibinfo {author} {\bibfnamefont {S.}~\bibnamefont {Jain}}, \bibinfo {author} {\bibfnamefont {Y.}~\bibnamefont {Ye}}, \bibinfo {author} {\bibfnamefont {S.~S.}\ \bibnamefont {Srinivasa}}, \bibinfo {author} {\bibfnamefont {M.}~\bibnamefont {Stilman}},\ and\ \bibinfo {author} {\bibfnamefont {C.~K.}\ \bibnamefont {Liu}},\ }\href {https://doi.org/10.21105/joss.00500} {\bibfield  {journal} {\bibinfo  {journal} {Journal of Open Source Software}\ }\textbf {\bibinfo {volume} {3}},\ \bibinfo {pages} {500} (\bibinfo {year} {2018})}\BibitemShut {NoStop}%
\bibitem [{\citenamefont {Coumans}\ and\ \citenamefont {Bai}(2017)}]{pybullet}%
  \BibitemOpen
  \bibfield  {author} {\bibinfo {author} {\bibfnamefont {E.}~\bibnamefont {Coumans}}\ and\ \bibinfo {author} {\bibfnamefont {Y.}~\bibnamefont {Bai}},\ }\href@noop {} {\bibinfo {title} {Pybullet, a python module for physics simulation in robotics, games and machine learning}} (\bibinfo {year} {2017})\BibitemShut {NoStop}%
\bibitem [{\citenamefont {Todorov}\ \emph {et~al.}(2012)\citenamefont {Todorov}, \citenamefont {Erez},\ and\ \citenamefont {Tassa}}]{mujoco}%
  \BibitemOpen
  \bibfield  {author} {\bibinfo {author} {\bibfnamefont {E.}~\bibnamefont {Todorov}}, \bibinfo {author} {\bibfnamefont {T.}~\bibnamefont {Erez}},\ and\ \bibinfo {author} {\bibfnamefont {Y.}~\bibnamefont {Tassa}},\ }in\ \href@noop {} {\emph {\bibinfo {booktitle} {Proceedings of the IEEE/RSJ International Conference on Intelligent Robots and Systems.}}}\ (\bibinfo {year} {2012})\ pp.\ \bibinfo {pages} {5026--5033}\BibitemShut {NoStop}%
\bibitem [{\citenamefont {Brockman}\ \emph {et~al.}(2016)\citenamefont {Brockman}, \citenamefont {Cheung}, \citenamefont {Pettersson}, \citenamefont {Schneider}, \citenamefont {Schulman}, \citenamefont {Tang},\ and\ \citenamefont {Zaremba}}]{brockman2016openai}%
  \BibitemOpen
  \bibfield  {author} {\bibinfo {author} {\bibfnamefont {G.}~\bibnamefont {Brockman}}, \bibinfo {author} {\bibfnamefont {V.}~\bibnamefont {Cheung}}, \bibinfo {author} {\bibfnamefont {L.}~\bibnamefont {Pettersson}}, \bibinfo {author} {\bibfnamefont {J.}~\bibnamefont {Schneider}}, \bibinfo {author} {\bibfnamefont {J.}~\bibnamefont {Schulman}}, \bibinfo {author} {\bibfnamefont {J.}~\bibnamefont {Tang}},\ and\ \bibinfo {author} {\bibfnamefont {W.}~\bibnamefont {Zaremba}},\ }\href@noop {} {\bibfield  {journal} {\bibinfo  {journal} {arXiv preprint arXiv:1606.01540}\ } (\bibinfo {year} {2016})}\BibitemShut {NoStop}%
\bibitem [{\citenamefont {Musgrave}\ \emph {et~al.}(2020)\citenamefont {Musgrave}, \citenamefont {Belongie},\ and\ \citenamefont {Lim}}]{stuck}%
  \BibitemOpen
  \bibfield  {author} {\bibinfo {author} {\bibfnamefont {K.}~\bibnamefont {Musgrave}}, \bibinfo {author} {\bibfnamefont {S.~J.}\ \bibnamefont {Belongie}},\ and\ \bibinfo {author} {\bibfnamefont {S.}~\bibnamefont {Lim}},\ }\href {https://arxiv.org/abs/2003.08505} {\bibfield  {journal} {\bibinfo  {journal} {CoRR}\ }\textbf {\bibinfo {volume} {abs/2003.08505}} (\bibinfo {year} {2020})},\ \Eprint {https://arxiv.org/abs/2003.08505} {arXiv:2003.08505} \BibitemShut {NoStop}%
\bibitem [{\citenamefont {Kapoor}\ and\ \citenamefont {Narayanan}(2022)}]{leakage}%
  \BibitemOpen
  \bibfield  {author} {\bibinfo {author} {\bibfnamefont {S.}~\bibnamefont {Kapoor}}\ and\ \bibinfo {author} {\bibfnamefont {A.}~\bibnamefont {Narayanan}},\ }\href {https://doi.org/10.48550/ARXIV.2207.07048} {\bibinfo {title} {Leakage and the reproducibility crisis in ml-based science}} (\bibinfo {year} {2022})\BibitemShut {NoStop}%
\bibitem [{\citenamefont {Whiting}\ \emph {et~al.}(2020)\citenamefont {Whiting}, \citenamefont {Campbell}, \citenamefont {Kang},\ and\ \citenamefont {Werner}}]{whiting2020meta}%
  \BibitemOpen
  \bibfield  {author} {\bibinfo {author} {\bibfnamefont {E.~B.}\ \bibnamefont {Whiting}}, \bibinfo {author} {\bibfnamefont {S.~D.}\ \bibnamefont {Campbell}}, \bibinfo {author} {\bibfnamefont {L.}~\bibnamefont {Kang}},\ and\ \bibinfo {author} {\bibfnamefont {D.~H.}\ \bibnamefont {Werner}},\ }\href@noop {} {\bibfield  {journal} {\bibinfo  {journal} {Optics Express}\ }\textbf {\bibinfo {volume} {28}},\ \bibinfo {pages} {24229} (\bibinfo {year} {2020})}\BibitemShut {NoStop}%
\bibitem [{\citenamefont {Tikhonravov}\ \emph {et~al.}(1996)\citenamefont {Tikhonravov}, \citenamefont {Trubetskov},\ and\ \citenamefont {DeBell}}]{tikhonravov1996application}%
  \BibitemOpen
  \bibfield  {author} {\bibinfo {author} {\bibfnamefont {A.~V.}\ \bibnamefont {Tikhonravov}}, \bibinfo {author} {\bibfnamefont {M.~K.}\ \bibnamefont {Trubetskov}},\ and\ \bibinfo {author} {\bibfnamefont {G.~W.}\ \bibnamefont {DeBell}},\ }\href@noop {} {\bibfield  {journal} {\bibinfo  {journal} {Applied optics}\ }\textbf {\bibinfo {volume} {35}},\ \bibinfo {pages} {5493} (\bibinfo {year} {1996})}\BibitemShut {NoStop}%
\bibitem [{\citenamefont {Cea}(1986)}]{ceaConceptionOptimaleOu1986}%
  \BibitemOpen
  \bibfield  {author} {\bibinfo {author} {\bibfnamefont {J.}~\bibnamefont {Cea}},\ }\href {https://doi.org/10.1051/m2an/1986200303711} {\bibfield  {journal} {\bibinfo  {journal} {ESAIM: Mathematical Modelling and Numerical Analysis}\ }\textbf {\bibinfo {volume} {20}},\ \bibinfo {pages} {371} (\bibinfo {year} {1986})}\BibitemShut {NoStop}%
\bibitem [{\citenamefont {Su}\ \emph {et~al.}(2020)\citenamefont {Su}, \citenamefont {Vercruysse}, \citenamefont {Skarda}, \citenamefont {Sapra}, \citenamefont {Petykiewicz},\ and\ \citenamefont {Vu{\v{c}}kovi{\'c}}}]{su2020nanophotonic}%
  \BibitemOpen
  \bibfield  {author} {\bibinfo {author} {\bibfnamefont {L.}~\bibnamefont {Su}}, \bibinfo {author} {\bibfnamefont {D.}~\bibnamefont {Vercruysse}}, \bibinfo {author} {\bibfnamefont {J.}~\bibnamefont {Skarda}}, \bibinfo {author} {\bibfnamefont {N.~V.}\ \bibnamefont {Sapra}}, \bibinfo {author} {\bibfnamefont {J.~A.}\ \bibnamefont {Petykiewicz}},\ and\ \bibinfo {author} {\bibfnamefont {J.}~\bibnamefont {Vu{\v{c}}kovi{\'c}}},\ }\href@noop {} {\bibfield  {journal} {\bibinfo  {journal} {Applied Physics Reviews}\ }\textbf {\bibinfo {volume} {7}} (\bibinfo {year} {2020})}\BibitemShut {NoStop}%
\bibitem [{\citenamefont {Wang}\ \emph {et~al.}(2003)\citenamefont {Wang}, \citenamefont {Wang},\ and\ \citenamefont {Guo}}]{wang2003level}%
  \BibitemOpen
  \bibfield  {author} {\bibinfo {author} {\bibfnamefont {M.~Y.}\ \bibnamefont {Wang}}, \bibinfo {author} {\bibfnamefont {X.}~\bibnamefont {Wang}},\ and\ \bibinfo {author} {\bibfnamefont {D.}~\bibnamefont {Guo}},\ }\href@noop {} {\bibfield  {journal} {\bibinfo  {journal} {Computer methods in applied mechanics and engineering}\ }\textbf {\bibinfo {volume} {192}},\ \bibinfo {pages} {227} (\bibinfo {year} {2003})}\BibitemShut {NoStop}%
\bibitem [{\citenamefont {Wang}\ \emph {et~al.}(2018)\citenamefont {Wang}, \citenamefont {Christiansen}, \citenamefont {Yu}, \citenamefont {M{\o}rk},\ and\ \citenamefont {Sigmund}}]{wang2018maximizing}%
  \BibitemOpen
  \bibfield  {author} {\bibinfo {author} {\bibfnamefont {F.}~\bibnamefont {Wang}}, \bibinfo {author} {\bibfnamefont {R.~E.}\ \bibnamefont {Christiansen}}, \bibinfo {author} {\bibfnamefont {Y.}~\bibnamefont {Yu}}, \bibinfo {author} {\bibfnamefont {J.}~\bibnamefont {M{\o}rk}},\ and\ \bibinfo {author} {\bibfnamefont {O.}~\bibnamefont {Sigmund}},\ }\href@noop {} {\bibfield  {journal} {\bibinfo  {journal} {arXiv preprint arXiv:1810.02417}\ } (\bibinfo {year} {2018})}\BibitemShut {NoStop}%
\bibitem [{\citenamefont {Martin}\ \emph {et~al.}(1995)\citenamefont {Martin}, \citenamefont {Rivory},\ and\ \citenamefont {Schoenauer}}]{martin1995synthesis}%
  \BibitemOpen
  \bibfield  {author} {\bibinfo {author} {\bibfnamefont {S.}~\bibnamefont {Martin}}, \bibinfo {author} {\bibfnamefont {J.}~\bibnamefont {Rivory}},\ and\ \bibinfo {author} {\bibfnamefont {M.}~\bibnamefont {Schoenauer}},\ }\href@noop {} {\bibfield  {journal} {\bibinfo  {journal} {applied Optics}\ }\textbf {\bibinfo {volume} {34}},\ \bibinfo {pages} {2247} (\bibinfo {year} {1995})}\BibitemShut {NoStop}%
\bibitem [{\citenamefont {Moreau}(2023)}]{moreauPyMoosh2023}%
  \BibitemOpen
  \bibfield  {author} {\bibinfo {author} {\bibfnamefont {A.}~\bibnamefont {Moreau}},\ }\href {https://github.com/AnMoreau/PyMoosh} {\bibinfo {title} {{{PyMoosh}}}} (\bibinfo {year} {2023})\BibitemShut {NoStop}%
\bibitem [{\citenamefont {Tikhonravov}(1993)}]{tikhonravov1993some}%
  \BibitemOpen
  \bibfield  {author} {\bibinfo {author} {\bibfnamefont {A.~V.}\ \bibnamefont {Tikhonravov}},\ }\href@noop {} {\bibfield  {journal} {\bibinfo  {journal} {Applied Optics}\ }\textbf {\bibinfo {volume} {32}},\ \bibinfo {pages} {5417} (\bibinfo {year} {1993})}\BibitemShut {NoStop}%
\bibitem [{\citenamefont {Br{\^u}l{\'e}}\ \emph {et~al.}(2022)\citenamefont {Br{\^u}l{\'e}}, \citenamefont {Wiecha}, \citenamefont {Cuche}, \citenamefont {Paillard},\ and\ \citenamefont {Des~Francs}}]{bruleMagneticElectricPurcell2022}%
  \BibitemOpen
  \bibfield  {author} {\bibinfo {author} {\bibfnamefont {Y.}~\bibnamefont {Br{\^u}l{\'e}}}, \bibinfo {author} {\bibfnamefont {P.}~\bibnamefont {Wiecha}}, \bibinfo {author} {\bibfnamefont {A.}~\bibnamefont {Cuche}}, \bibinfo {author} {\bibfnamefont {V.}~\bibnamefont {Paillard}},\ and\ \bibinfo {author} {\bibfnamefont {G.~C.}\ \bibnamefont {Des~Francs}},\ }\href@noop {} {\bibfield  {journal} {\bibinfo  {journal} {Optics Express}\ }\textbf {\bibinfo {volume} {30}},\ \bibinfo {pages} {20360} (\bibinfo {year} {2022})}\BibitemShut {NoStop}%
\bibitem [{\citenamefont {Yang}\ \emph {et~al.}(2023)\citenamefont {Yang}, \citenamefont {Guidry}, \citenamefont {Lukin}, \citenamefont {Yang},\ and\ \citenamefont {Vu{\v{c}}kovi{\'c}}}]{yang2023inverse}%
  \BibitemOpen
  \bibfield  {author} {\bibinfo {author} {\bibfnamefont {J.}~\bibnamefont {Yang}}, \bibinfo {author} {\bibfnamefont {M.~A.}\ \bibnamefont {Guidry}}, \bibinfo {author} {\bibfnamefont {D.~M.}\ \bibnamefont {Lukin}}, \bibinfo {author} {\bibfnamefont {K.}~\bibnamefont {Yang}},\ and\ \bibinfo {author} {\bibfnamefont {J.}~\bibnamefont {Vu{\v{c}}kovi{\'c}}},\ }\href@noop {} {\bibfield  {journal} {\bibinfo  {journal} {arXiv preprint arXiv:2303.17079}\ } (\bibinfo {year} {2023})}\BibitemShut {NoStop}%
\bibitem [{\citenamefont {Smaali}\ \emph {et~al.}(2021)\citenamefont {Smaali}, \citenamefont {Taliercio}, \citenamefont {Moreau},\ and\ \citenamefont {Centeno}}]{smaali2021reshaping}%
  \BibitemOpen
  \bibfield  {author} {\bibinfo {author} {\bibfnamefont {R.}~\bibnamefont {Smaali}}, \bibinfo {author} {\bibfnamefont {T.}~\bibnamefont {Taliercio}}, \bibinfo {author} {\bibfnamefont {A.}~\bibnamefont {Moreau}},\ and\ \bibinfo {author} {\bibfnamefont {E.}~\bibnamefont {Centeno}},\ }\href@noop {} {\bibfield  {journal} {\bibinfo  {journal} {Applied Physics Letters}\ }\textbf {\bibinfo {volume} {119}} (\bibinfo {year} {2021})}\BibitemShut {NoStop}%
\bibitem [{\citenamefont {Centeno}\ \emph {et~al.}(2021)\citenamefont {Centeno}, \citenamefont {Alvear-Cabez{\'o}n}, \citenamefont {Smaali}, \citenamefont {Moreau},\ and\ \citenamefont {Taliercio}}]{centeno2021inverse}%
  \BibitemOpen
  \bibfield  {author} {\bibinfo {author} {\bibfnamefont {E.}~\bibnamefont {Centeno}}, \bibinfo {author} {\bibfnamefont {E.}~\bibnamefont {Alvear-Cabez{\'o}n}}, \bibinfo {author} {\bibfnamefont {R.}~\bibnamefont {Smaali}}, \bibinfo {author} {\bibfnamefont {A.}~\bibnamefont {Moreau}},\ and\ \bibinfo {author} {\bibfnamefont {T.}~\bibnamefont {Taliercio}},\ }\href@noop {} {\bibfield  {journal} {\bibinfo  {journal} {Semiconductor Science and Technology}\ }\textbf {\bibinfo {volume} {36}},\ \bibinfo {pages} {085014} (\bibinfo {year} {2021})}\BibitemShut {NoStop}%
\bibitem [{\citenamefont {Moreau}\ \emph {et~al.}(2007)\citenamefont {Moreau}, \citenamefont {Lafarge}, \citenamefont {Laurent}, \citenamefont {Edee},\ and\ \citenamefont {Granet}}]{moreau2007enhanced}%
  \BibitemOpen
  \bibfield  {author} {\bibinfo {author} {\bibfnamefont {A.}~\bibnamefont {Moreau}}, \bibinfo {author} {\bibfnamefont {C.}~\bibnamefont {Lafarge}}, \bibinfo {author} {\bibfnamefont {N.}~\bibnamefont {Laurent}}, \bibinfo {author} {\bibfnamefont {K.}~\bibnamefont {Edee}},\ and\ \bibinfo {author} {\bibfnamefont {G.}~\bibnamefont {Granet}},\ }\href {https://doi.org/10.1088/1464-4258/9/2/008} {\bibfield  {journal} {\bibinfo  {journal} {Journal of Optics A: Pure and Applied Optics}\ }\textbf {\bibinfo {volume} {9}},\ \bibinfo {pages} {165} (\bibinfo {year} {2007})}\BibitemShut {NoStop}%
\bibitem [{\citenamefont {Santbergen}\ \emph {et~al.}(2010)\citenamefont {Santbergen}, \citenamefont {Goud}, \citenamefont {Zeman}, \citenamefont {van Roosmalen},\ and\ \citenamefont {van Zolingen}}]{santbergen2010am1}%
  \BibitemOpen
  \bibfield  {author} {\bibinfo {author} {\bibfnamefont {R.}~\bibnamefont {Santbergen}}, \bibinfo {author} {\bibfnamefont {J.}~\bibnamefont {Goud}}, \bibinfo {author} {\bibfnamefont {M.}~\bibnamefont {Zeman}}, \bibinfo {author} {\bibfnamefont {J.}~\bibnamefont {van Roosmalen}},\ and\ \bibinfo {author} {\bibfnamefont {R.~C.}\ \bibnamefont {van Zolingen}},\ }\href@noop {} {\bibfield  {journal} {\bibinfo  {journal} {Solar energy materials and solar cells}\ }\textbf {\bibinfo {volume} {94}},\ \bibinfo {pages} {715} (\bibinfo {year} {2010})}\BibitemShut {NoStop}%
\bibitem [{\citenamefont {Wiecha}\ \emph {et~al.}(2019)\citenamefont {Wiecha}, \citenamefont {Majorel}, \citenamefont {Girard}, \citenamefont {Cuche}, \citenamefont {Paillard}, \citenamefont {Muskens}, \citenamefont {Arbouet},\ and\ \citenamefont {Arbouet}}]{wiechaDesignPlasmonicDirectional2019}%
  \BibitemOpen
  \bibfield  {author} {\bibinfo {author} {\bibfnamefont {P.~R.}\ \bibnamefont {Wiecha}}, \bibinfo {author} {\bibfnamefont {C.}~\bibnamefont {Majorel}}, \bibinfo {author} {\bibfnamefont {C.}~\bibnamefont {Girard}}, \bibinfo {author} {\bibfnamefont {A.}~\bibnamefont {Cuche}}, \bibinfo {author} {\bibfnamefont {V.}~\bibnamefont {Paillard}}, \bibinfo {author} {\bibfnamefont {O.~L.}\ \bibnamefont {Muskens}}, \bibinfo {author} {\bibfnamefont {A.}~\bibnamefont {Arbouet}},\ and\ \bibinfo {author} {\bibfnamefont {A.}~\bibnamefont {Arbouet}},\ }\href {https://doi.org/10.1364/OE.27.029069} {\bibfield  {journal} {\bibinfo  {journal} {Optics Express}\ }\textbf {\bibinfo {volume} {27}},\ \bibinfo {pages} {29069} (\bibinfo {year} {2019})}\BibitemShut {NoStop}%
\bibitem [{\citenamefont {Hansen}\ \emph {et~al.}(2019)\citenamefont {Hansen}, \citenamefont {Akimoto},\ and\ \citenamefont {Baudis}}]{pycma}%
  \BibitemOpen
  \bibfield  {author} {\bibinfo {author} {\bibfnamefont {N.}~\bibnamefont {Hansen}}, \bibinfo {author} {\bibfnamefont {Y.}~\bibnamefont {Akimoto}},\ and\ \bibinfo {author} {\bibfnamefont {P.}~\bibnamefont {Baudis}},\ }\href {https://doi.org/10.5281/zenodo.2559634} {\bibinfo {title} {{CMA-ES/pycma} on {G}ithub}},\ \bibinfo {howpublished} {Zenodo, DOI:10.5281/zenodo.2559634} (\bibinfo {year} {2019})\BibitemShut {NoStop}%
\bibitem [{\citenamefont {FacebookResearch}(2020)}]{ax}%
  \BibitemOpen
  \bibfield  {author} {\bibinfo {author} {\bibnamefont {FacebookResearch}},\ }\href@noop {} {\bibinfo {title} {Ax - adaptive experimentation}},\ \bibinfo {howpublished} {\url{ ax.dev}} (\bibinfo {year} {2020})\BibitemShut {NoStop}%
\bibitem [{\citenamefont {Bergstra}\ \emph {et~al.}(2015)\citenamefont {Bergstra}, \citenamefont {Komer}, \citenamefont {Eliasmith}, \citenamefont {Yamins},\ and\ \citenamefont {Cox}}]{hyperopt}%
  \BibitemOpen
  \bibfield  {author} {\bibinfo {author} {\bibfnamefont {J.}~\bibnamefont {Bergstra}}, \bibinfo {author} {\bibfnamefont {B.}~\bibnamefont {Komer}}, \bibinfo {author} {\bibfnamefont {C.}~\bibnamefont {Eliasmith}}, \bibinfo {author} {\bibfnamefont {D.}~\bibnamefont {Yamins}},\ and\ \bibinfo {author} {\bibfnamefont {D.~D.}\ \bibnamefont {Cox}},\ }\href@noop {} {\bibfield  {journal} {\bibinfo  {journal} {Computational Science and Discovery}\ }\textbf {\bibinfo {volume} {8}},\ \bibinfo {pages} {014008} (\bibinfo {year} {2015})}\BibitemShut {NoStop}%
\bibitem [{\citenamefont {Hutter}\ \emph {et~al.}(2011)\citenamefont {Hutter}, \citenamefont {Hoos},\ and\ \citenamefont {Leyton-Brown}}]{smac}%
  \BibitemOpen
  \bibfield  {author} {\bibinfo {author} {\bibfnamefont {F.}~\bibnamefont {Hutter}}, \bibinfo {author} {\bibfnamefont {H.~H.}\ \bibnamefont {Hoos}},\ and\ \bibinfo {author} {\bibfnamefont {K.}~\bibnamefont {Leyton-Brown}},\ }in\ \href {http://dblp.uni-trier.de/db/conf/lion/lion2011.html\#HutterHL11} {\emph {\bibinfo {booktitle} {LION}}},\ \bibinfo {series} {Lecture Notes in Computer Science}, Vol.\ \bibinfo {volume} {6683},\ \bibinfo {editor} {edited by\ \bibinfo {editor} {\bibfnamefont {C.~A.~C.}\ \bibnamefont {Coello}}}\ (\bibinfo  {publisher} {Springer},\ \bibinfo {year} {2011})\ pp.\ \bibinfo {pages} {507--523}\BibitemShut {NoStop}%
\bibitem [{\citenamefont {Johnson}(1994)}]{nlopt}%
  \BibitemOpen
  \bibfield  {author} {\bibinfo {author} {\bibfnamefont {S.~G.}\ \bibnamefont {Johnson}},\ }\href {http://github.com/stevengj/nlopt} {\bibinfo {title} {The nlopt nonlinear-optimization package}} (\bibinfo {year} {1994})\BibitemShut {NoStop}%
\bibitem [{\citenamefont {Liu}\ and\ \citenamefont {Nocedal}(1989)}]{lbfgs}%
  \BibitemOpen
  \bibfield  {author} {\bibinfo {author} {\bibfnamefont {D.~C.}\ \bibnamefont {Liu}}\ and\ \bibinfo {author} {\bibfnamefont {J.}~\bibnamefont {Nocedal}},\ }\href {https://doi.org/10.1007/BF01589116} {\bibfield  {journal} {\bibinfo  {journal} {Math. Program.}\ }\textbf {\bibinfo {volume} {45}},\ \bibinfo {pages} {503} (\bibinfo {year} {1989})}\BibitemShut {NoStop}%
\bibitem [{\citenamefont {Allaire}(2015)}]{allaire2015review}%
  \BibitemOpen
  \bibfield  {author} {\bibinfo {author} {\bibfnamefont {G.}~\bibnamefont {Allaire}},\ }\href@noop {} {\bibfield  {journal} {\bibinfo  {journal} {Ing{\'e}nieurs de l'Automobile}\ }\textbf {\bibinfo {volume} {836}},\ \bibinfo {pages} {33} (\bibinfo {year} {2015})}\BibitemShut {NoStop}%
\bibitem [{\citenamefont {Maclaurin}\ \emph {et~al.}(2015)\citenamefont {Maclaurin}, \citenamefont {Duvenaud},\ and\ \citenamefont {Adams}}]{maclaurin2015autograd}%
  \BibitemOpen
  \bibfield  {author} {\bibinfo {author} {\bibfnamefont {D.}~\bibnamefont {Maclaurin}}, \bibinfo {author} {\bibfnamefont {D.}~\bibnamefont {Duvenaud}},\ and\ \bibinfo {author} {\bibfnamefont {R.~P.}\ \bibnamefont {Adams}},\ }in\ \href@noop {} {\emph {\bibinfo {booktitle} {ICML 2015 AutoML workshop}}},\ Vol.\ \bibinfo {volume} {238}\ (\bibinfo {year} {2015})\BibitemShut {NoStop}%
\bibitem [{\citenamefont {Janikow}\ \emph {et~al.}(1991)\citenamefont {Janikow}, \citenamefont {Michalewicz} \emph {et~al.}}]{janikow1991experimental}%
  \BibitemOpen
  \bibfield  {author} {\bibinfo {author} {\bibfnamefont {C.~Z.}\ \bibnamefont {Janikow}}, \bibinfo {author} {\bibfnamefont {Z.}~\bibnamefont {Michalewicz}}, \emph {et~al.},\ }in\ \href@noop {} {\emph {\bibinfo {booktitle} {ICGA}}},\ Vol.\ \bibinfo {volume} {1991}\ (\bibinfo {year} {1991})\ pp.\ \bibinfo {pages} {31--36}\BibitemShut {NoStop}%
\bibitem [{\citenamefont {Herrera}\ \emph {et~al.}(1998)\citenamefont {Herrera}, \citenamefont {Lozano},\ and\ \citenamefont {Verdegay}}]{herrera1998tackling}%
  \BibitemOpen
  \bibfield  {author} {\bibinfo {author} {\bibfnamefont {F.}~\bibnamefont {Herrera}}, \bibinfo {author} {\bibfnamefont {M.}~\bibnamefont {Lozano}},\ and\ \bibinfo {author} {\bibfnamefont {J.~L.}\ \bibnamefont {Verdegay}},\ }\href@noop {} {\bibfield  {journal} {\bibinfo  {journal} {Artificial intelligence review}\ }\textbf {\bibinfo {volume} {12}},\ \bibinfo {pages} {265} (\bibinfo {year} {1998})}\BibitemShut {NoStop}%
\bibitem [{\citenamefont {Storn}\ and\ \citenamefont {Price}(1997)}]{de}%
  \BibitemOpen
  \bibfield  {author} {\bibinfo {author} {\bibfnamefont {R.}~\bibnamefont {Storn}}\ and\ \bibinfo {author} {\bibfnamefont {K.}~\bibnamefont {Price}},\ }\href@noop {} {\bibfield  {journal} {\bibinfo  {journal} {J. of Global Optimization}\ }\textbf {\bibinfo {volume} {11}},\ \bibinfo {pages} {341} (\bibinfo {year} {1997})}\BibitemShut {NoStop}%
\bibitem [{\citenamefont {{Rahnamayan}}\ \emph {et~al.}(2007)\citenamefont {{Rahnamayan}}, \citenamefont {{Tizhoosh}},\ and\ \citenamefont {{Salama}}}]{quasiopposite}%
  \BibitemOpen
  \bibfield  {author} {\bibinfo {author} {\bibfnamefont {S.}~\bibnamefont {{Rahnamayan}}}, \bibinfo {author} {\bibfnamefont {H.~R.}\ \bibnamefont {{Tizhoosh}}},\ and\ \bibinfo {author} {\bibfnamefont {M.~M.~A.}\ \bibnamefont {{Salama}}},\ }in\ \href@noop {} {\emph {\bibinfo {booktitle} {2007 IEEE Congress on Evolutionary Computation}}}\ (\bibinfo {year} {2007})\ pp.\ \bibinfo {pages} {2229--2236}\BibitemShut {NoStop}%
\bibitem [{\citenamefont {Schevenels}\ \emph {et~al.}(2011)\citenamefont {Schevenels}, \citenamefont {Lazarov},\ and\ \citenamefont {Sigmund}}]{robustoptim}%
  \BibitemOpen
  \bibfield  {author} {\bibinfo {author} {\bibfnamefont {M.}~\bibnamefont {Schevenels}}, \bibinfo {author} {\bibfnamefont {B.}~\bibnamefont {Lazarov}},\ and\ \bibinfo {author} {\bibfnamefont {O.}~\bibnamefont {Sigmund}},\ }\href {https://doi.org/https://doi.org/10.1016/j.cma.2011.08.006} {\bibfield  {journal} {\bibinfo  {journal} {Computer Methods in Applied Mechanics and Engineering}\ }\textbf {\bibinfo {volume} {200}},\ \bibinfo {pages} {3613} (\bibinfo {year} {2011})}\BibitemShut {NoStop}%
\bibitem [{\citenamefont {Beyer}\ and\ \citenamefont {Sendhoff}(2007)}]{robustoptim2}%
  \BibitemOpen
  \bibfield  {author} {\bibinfo {author} {\bibfnamefont {H.-G.}\ \bibnamefont {Beyer}}\ and\ \bibinfo {author} {\bibfnamefont {B.}~\bibnamefont {Sendhoff}},\ }\href {https://doi.org/https://doi.org/10.1016/j.cma.2007.03.003} {\bibfield  {journal} {\bibinfo  {journal} {Computer Methods in Applied Mechanics and Engineering}\ }\textbf {\bibinfo {volume} {196}},\ \bibinfo {pages} {3190} (\bibinfo {year} {2007})}\BibitemShut {NoStop}%
\bibitem [{\citenamefont {Rapin}\ \emph {et~al.}(2020)\citenamefont {Rapin}, \citenamefont {Bennet}, \citenamefont {Centeno}, \citenamefont {Haziza}, \citenamefont {Moreau},\ and\ \citenamefont {Teytaud}}]{rapin2020open}%
  \BibitemOpen
  \bibfield  {author} {\bibinfo {author} {\bibfnamefont {J.}~\bibnamefont {Rapin}}, \bibinfo {author} {\bibfnamefont {P.}~\bibnamefont {Bennet}}, \bibinfo {author} {\bibfnamefont {E.}~\bibnamefont {Centeno}}, \bibinfo {author} {\bibfnamefont {D.}~\bibnamefont {Haziza}}, \bibinfo {author} {\bibfnamefont {A.}~\bibnamefont {Moreau}},\ and\ \bibinfo {author} {\bibfnamefont {O.}~\bibnamefont {Teytaud}},\ }in\ \href@noop {} {\emph {\bibinfo {booktitle} {Proceedings of the 2020 Genetic and Evolutionary Computation Conference Companion}}}\ (\bibinfo {year} {2020})\ pp.\ \bibinfo {pages} {1599--1607}\BibitemShut {NoStop}%
\bibitem [{\citenamefont {Kennedy}\ and\ \citenamefont {Eberhart}(1995)}]{pso}%
  \BibitemOpen
  \bibfield  {author} {\bibinfo {author} {\bibfnamefont {J.}~\bibnamefont {Kennedy}}\ and\ \bibinfo {author} {\bibfnamefont {R.~C.}\ \bibnamefont {Eberhart}},\ }in\ \href@noop {} {\emph {\bibinfo {booktitle} {Proceedings of the IEEE International Conference on Neural Networks}}}\ (\bibinfo {year} {1995})\ pp.\ \bibinfo {pages} {1942--1948}\BibitemShut {NoStop}%
\bibitem [{\citenamefont {Hansen}\ and\ \citenamefont {Ostermeier}(2003)}]{CMA}%
  \BibitemOpen
  \bibfield  {author} {\bibinfo {author} {\bibfnamefont {N.}~\bibnamefont {Hansen}}\ and\ \bibinfo {author} {\bibfnamefont {A.}~\bibnamefont {Ostermeier}},\ }\href@noop {} {\bibfield  {journal} {\bibinfo  {journal} {Evolutionary Computation}\ }\textbf {\bibinfo {volume} {11}} (\bibinfo {year} {2003})}\BibitemShut {NoStop}%
\bibitem [{\citenamefont {Rechenberg}(1973)}]{rechenberg73}%
  \BibitemOpen
  \bibfield  {author} {\bibinfo {author} {\bibfnamefont {I.}~\bibnamefont {Rechenberg}},\ }\href@noop {} {\emph {\bibinfo {title} {Evolutionsstrategie {O}ptimierung technischer {S}ysteme nach {P}rinzipien der biologischen {E}volution}}}\ (\bibinfo  {publisher} {Friedrich Frommann Verlag},\ \bibinfo {address} {Stuttgart-Bad Cannstatt},\ \bibinfo {year} {1973})\BibitemShut {NoStop}%
\bibitem [{\citenamefont {Jones}\ \emph {et~al.}(1998)\citenamefont {Jones}, \citenamefont {Schonlau},\ and\ \citenamefont {Welch}}]{ego}%
  \BibitemOpen
  \bibfield  {author} {\bibinfo {author} {\bibfnamefont {D.~R.}\ \bibnamefont {Jones}}, \bibinfo {author} {\bibfnamefont {M.}~\bibnamefont {Schonlau}},\ and\ \bibinfo {author} {\bibfnamefont {W.~J.}\ \bibnamefont {Welch}},\ }\href@noop {} {\bibfield  {journal} {\bibinfo  {journal} {Journal of Global Optimization}\ }\textbf {\bibinfo {volume} {13}},\ \bibinfo {pages} {455} (\bibinfo {year} {1998})}\BibitemShut {NoStop}%
\bibitem [{\citenamefont {Elsawy}\ \emph {et~al.}(2021)\citenamefont {Elsawy}, \citenamefont {Gourdin}, \citenamefont {Binois}, \citenamefont {Duvigneau}, \citenamefont {Felbacq}, \citenamefont {Khadir}, \citenamefont {Genevet},\ and\ \citenamefont {Lanteri}}]{elsawy2021multiobjective}%
  \BibitemOpen
  \bibfield  {author} {\bibinfo {author} {\bibfnamefont {M.~M.}\ \bibnamefont {Elsawy}}, \bibinfo {author} {\bibfnamefont {A.}~\bibnamefont {Gourdin}}, \bibinfo {author} {\bibfnamefont {M.}~\bibnamefont {Binois}}, \bibinfo {author} {\bibfnamefont {R.}~\bibnamefont {Duvigneau}}, \bibinfo {author} {\bibfnamefont {D.}~\bibnamefont {Felbacq}}, \bibinfo {author} {\bibfnamefont {S.}~\bibnamefont {Khadir}}, \bibinfo {author} {\bibfnamefont {P.}~\bibnamefont {Genevet}},\ and\ \bibinfo {author} {\bibfnamefont {S.}~\bibnamefont {Lanteri}},\ }\href@noop {} {\bibfield  {journal} {\bibinfo  {journal} {ACS photonics}\ }\textbf {\bibinfo {volume} {8}},\ \bibinfo {pages} {2498} (\bibinfo {year} {2021})}\BibitemShut {NoStop}%
\bibitem [{\citenamefont {Hutter}\ \emph {et~al.}(2013)\citenamefont {Hutter}, \citenamefont {Hoos},\ and\ \citenamefont {Leyton-Brown}}]{bbobsmac}%
  \BibitemOpen
  \bibfield  {author} {\bibinfo {author} {\bibfnamefont {F.}~\bibnamefont {Hutter}}, \bibinfo {author} {\bibfnamefont {H.}~\bibnamefont {Hoos}},\ and\ \bibinfo {author} {\bibfnamefont {K.}~\bibnamefont {Leyton-Brown}},\ }in\ \href {https://doi.org/10.1145/2464576.2501592} {\emph {\bibinfo {booktitle} {Proceedings of the 15th Annual Conference Companion on Genetic and Evolutionary Computation}}},\ \bibinfo {series and number} {GECCO '13 Companion}\ (\bibinfo  {publisher} {Association for Computing Machinery},\ \bibinfo {address} {New York, NY, USA},\ \bibinfo {year} {2013})\ p.\ \bibinfo {pages} {1209–1216}\BibitemShut {NoStop}%
\bibitem [{\citenamefont {Nelder}\ and\ \citenamefont {Mead}(1965)}]{NM}%
  \BibitemOpen
  \bibfield  {author} {\bibinfo {author} {\bibfnamefont {J.~A.}\ \bibnamefont {Nelder}}\ and\ \bibinfo {author} {\bibfnamefont {R.}~\bibnamefont {Mead}},\ }\href@noop {} {\bibfield  {journal} {\bibinfo  {journal} {Computer Journal}\ }\textbf {\bibinfo {volume} {7}},\ \bibinfo {pages} {308} (\bibinfo {year} {1965})}\BibitemShut {NoStop}%
\bibitem [{\citenamefont {Liu}\ \emph {et~al.}(2020)\citenamefont {Liu}, \citenamefont {Moreau}, \citenamefont {Preuss}, \citenamefont {Rapin}, \citenamefont {Roziere}, \citenamefont {Teytaud},\ and\ \citenamefont {Teytaud}}]{versatile}%
  \BibitemOpen
  \bibfield  {author} {\bibinfo {author} {\bibfnamefont {J.}~\bibnamefont {Liu}}, \bibinfo {author} {\bibfnamefont {A.}~\bibnamefont {Moreau}}, \bibinfo {author} {\bibfnamefont {M.}~\bibnamefont {Preuss}}, \bibinfo {author} {\bibfnamefont {J.}~\bibnamefont {Rapin}}, \bibinfo {author} {\bibfnamefont {B.}~\bibnamefont {Roziere}}, \bibinfo {author} {\bibfnamefont {F.}~\bibnamefont {Teytaud}},\ and\ \bibinfo {author} {\bibfnamefont {O.}~\bibnamefont {Teytaud}},\ }in\ \href@noop {} {\emph {\bibinfo {booktitle} {Proceedings of the 2020 Genetic and Evolutionary Computation Conference}}},\ \bibinfo {series and number} {GECCO '20}\ (\bibinfo {year} {2020})\ p.\ \bibinfo {pages} {620–628}\BibitemShut {NoStop}%
\bibitem [{\citenamefont {Zoph}\ and\ \citenamefont {Le}(2017)}]{NAS}%
  \BibitemOpen
  \bibfield  {author} {\bibinfo {author} {\bibfnamefont {B.}~\bibnamefont {Zoph}}\ and\ \bibinfo {author} {\bibfnamefont {Q.~V.}\ \bibnamefont {Le}},\ }in\ \href {https://openreview.net/forum?id=r1Ue8Hcxg} {\emph {\bibinfo {booktitle} {5th International Conference on Learning Representations, {ICLR} 2017, Toulon, France, April 24-26, 2017, Conference Track Proceedings}}}\ (\bibinfo  {publisher} {OpenReview.net},\ \bibinfo {year} {2017})\BibitemShut {NoStop}%
\bibitem [{\citenamefont {Real}\ \emph {et~al.}(2017)\citenamefont {Real}, \citenamefont {Moore}, \citenamefont {Selle}, \citenamefont {Saxena}, \citenamefont {Suematsu}, \citenamefont {Le},\ and\ \citenamefont {Kurakin}}]{evo}%
  \BibitemOpen
  \bibfield  {author} {\bibinfo {author} {\bibfnamefont {E.}~\bibnamefont {Real}}, \bibinfo {author} {\bibfnamefont {S.}~\bibnamefont {Moore}}, \bibinfo {author} {\bibfnamefont {A.}~\bibnamefont {Selle}}, \bibinfo {author} {\bibfnamefont {S.}~\bibnamefont {Saxena}}, \bibinfo {author} {\bibfnamefont {Y.~L.}\ \bibnamefont {Suematsu}}, \bibinfo {author} {\bibfnamefont {Q.~V.}\ \bibnamefont {Le}},\ and\ \bibinfo {author} {\bibfnamefont {A.}~\bibnamefont {Kurakin}},\ }\href {http://arxiv.org/abs/1703.01041} {\bibfield  {journal} {\bibinfo  {journal} {CoRR}\ }\textbf {\bibinfo {volume} {abs/1703.01041}} (\bibinfo {year} {2017})},\ \Eprint {https://arxiv.org/abs/1703.01041} {arXiv:1703.01041} \BibitemShut {NoStop}%
\bibitem [{\citenamefont {Cheng}\ \emph {et~al.}(2023)\citenamefont {Cheng}, \citenamefont {Kahng}, \citenamefont {Kundu}, \citenamefont {Wang},\ and\ \citenamefont {Wang}}]{omcd}%
  \BibitemOpen
  \bibfield  {author} {\bibinfo {author} {\bibfnamefont {C.-K.}\ \bibnamefont {Cheng}}, \bibinfo {author} {\bibfnamefont {A.~B.}\ \bibnamefont {Kahng}}, \bibinfo {author} {\bibfnamefont {S.}~\bibnamefont {Kundu}}, \bibinfo {author} {\bibfnamefont {Y.}~\bibnamefont {Wang}},\ and\ \bibinfo {author} {\bibfnamefont {Z.}~\bibnamefont {Wang}},\ }in\ \href {https://doi.org/10.1145/3569052.3578926} {\emph {\bibinfo {booktitle} {Proceedings of the 2023 International Symposium on Physical Design}}},\ \bibinfo {series and number} {ISPD '23}\ (\bibinfo  {publisher} {Association for Computing Machinery},\ \bibinfo {address} {New York, NY, USA},\ \bibinfo {year} {2023})\ p.\ \bibinfo {pages} {158–166}\BibitemShut {NoStop}%
\bibitem [{\citenamefont {Langevin}\ \emph {et~al.}(2023)\citenamefont {Langevin}, \citenamefont {Bennet}, \citenamefont {Khaireh-Walieh}, \citenamefont {Wiecha}, \citenamefont {Teytaud},\ and\ \citenamefont {Moreau}}]{langevin2023pymoosh}%
  \BibitemOpen
  \bibfield  {author} {\bibinfo {author} {\bibfnamefont {D.}~\bibnamefont {Langevin}}, \bibinfo {author} {\bibfnamefont {P.}~\bibnamefont {Bennet}}, \bibinfo {author} {\bibfnamefont {A.}~\bibnamefont {Khaireh-Walieh}}, \bibinfo {author} {\bibfnamefont {P.}~\bibnamefont {Wiecha}}, \bibinfo {author} {\bibfnamefont {O.}~\bibnamefont {Teytaud}},\ and\ \bibinfo {author} {\bibfnamefont {A.}~\bibnamefont {Moreau}},\ }\href@noop {} {\bibinfo {title} {Pymoosh : a comprehensive numerical toolkit for computing the optical properties of multilayered structures}} (\bibinfo {year} {2023}),\ \Eprint {https://arxiv.org/abs/2309.00654} {arXiv:2309.00654 [physics.comp-ph]} \BibitemShut {NoStop}%
\bibitem [{\citenamefont {Lalanne}\ and\ \citenamefont {Morris}(1996)}]{lalanne1996highly}%
  \BibitemOpen
  \bibfield  {author} {\bibinfo {author} {\bibfnamefont {P.}~\bibnamefont {Lalanne}}\ and\ \bibinfo {author} {\bibfnamefont {G.~M.}\ \bibnamefont {Morris}},\ }\href@noop {} {\bibfield  {journal} {\bibinfo  {journal} {JOSA A}\ }\textbf {\bibinfo {volume} {13}},\ \bibinfo {pages} {779} (\bibinfo {year} {1996})}\BibitemShut {NoStop}%
\bibitem [{\citenamefont {Granet}\ and\ \citenamefont {Guizal}(1996)}]{granet1996efficient}%
  \BibitemOpen
  \bibfield  {author} {\bibinfo {author} {\bibfnamefont {G.}~\bibnamefont {Granet}}\ and\ \bibinfo {author} {\bibfnamefont {B.}~\bibnamefont {Guizal}},\ }\href@noop {} {\bibfield  {journal} {\bibinfo  {journal} {JOSA A}\ }\textbf {\bibinfo {volume} {13}},\ \bibinfo {pages} {1019} (\bibinfo {year} {1996})}\BibitemShut {NoStop}%
\bibitem [{\citenamefont {Wiecha}(2018)}]{wiecha_pygdm_2018}%
  \BibitemOpen
  \bibfield  {author} {\bibinfo {author} {\bibfnamefont {P.~R.}\ \bibnamefont {Wiecha}},\ }\href {https://doi.org/10.1016/j.cpc.2018.06.017} {\bibfield  {journal} {\bibinfo  {journal} {Computer Physics Communications}\ }\textbf {\bibinfo {volume} {233}},\ \bibinfo {pages} {167} (\bibinfo {year} {2018})},\ \Eprint {https://arxiv.org/abs/1802.04071} {1802.04071} \BibitemShut {NoStop}%
\bibitem [{\citenamefont {Wiecha}\ \emph {et~al.}(2022)\citenamefont {Wiecha}, \citenamefont {Majorel}, \citenamefont {Arbouet}, \citenamefont {Patoux}, \citenamefont {Br{\^u}l{\'e}}, \citenamefont {des Francs},\ and\ \citenamefont {Girard}}]{wiechaPyGDMNewFunctionalities2022}%
  \BibitemOpen
  \bibfield  {author} {\bibinfo {author} {\bibfnamefont {P.~R.}\ \bibnamefont {Wiecha}}, \bibinfo {author} {\bibfnamefont {C.}~\bibnamefont {Majorel}}, \bibinfo {author} {\bibfnamefont {A.}~\bibnamefont {Arbouet}}, \bibinfo {author} {\bibfnamefont {A.}~\bibnamefont {Patoux}}, \bibinfo {author} {\bibfnamefont {Y.}~\bibnamefont {Br{\^u}l{\'e}}}, \bibinfo {author} {\bibfnamefont {G.~C.}\ \bibnamefont {des Francs}},\ and\ \bibinfo {author} {\bibfnamefont {C.}~\bibnamefont {Girard}},\ }\href {https://doi.org/10.1016/j.cpc.2021.108142} {\bibfield  {journal} {\bibinfo  {journal} {Computer Physics Communications}\ }\textbf {\bibinfo {volume} {270}},\ \bibinfo {pages} {108142} (\bibinfo {year} {2022})},\ \Eprint {https://arxiv.org/abs/2105.04587} {2105.04587} \BibitemShut {NoStop}%
\bibitem [{\citenamefont {Schneider}\ \emph {et~al.}(2019)\citenamefont {Schneider}, \citenamefont {Garcia~Santiago}, \citenamefont {Soltwisch}, \citenamefont {Hammerschmidt}, \citenamefont {Burger},\ and\ \citenamefont {Rockstuhl}}]{schneider2019benchmarking}%
  \BibitemOpen
  \bibfield  {author} {\bibinfo {author} {\bibfnamefont {P.-I.}\ \bibnamefont {Schneider}}, \bibinfo {author} {\bibfnamefont {X.}~\bibnamefont {Garcia~Santiago}}, \bibinfo {author} {\bibfnamefont {V.}~\bibnamefont {Soltwisch}}, \bibinfo {author} {\bibfnamefont {M.}~\bibnamefont {Hammerschmidt}}, \bibinfo {author} {\bibfnamefont {S.}~\bibnamefont {Burger}},\ and\ \bibinfo {author} {\bibfnamefont {C.}~\bibnamefont {Rockstuhl}},\ }\href@noop {} {\bibfield  {journal} {\bibinfo  {journal} {ACS Photonics}\ }\textbf {\bibinfo {volume} {6}},\ \bibinfo {pages} {2726} (\bibinfo {year} {2019})}\BibitemShut {NoStop}%
\bibitem [{\citenamefont {Meunier}\ \emph {et~al.}(2021)\citenamefont {Meunier}, \citenamefont {Rakotoarison}, \citenamefont {Wong}, \citenamefont {Roziere}, \citenamefont {Rapin}, \citenamefont {Teytaud}, \citenamefont {Moreau},\ and\ \citenamefont {Doerr}}]{meunier2021black}%
  \BibitemOpen
  \bibfield  {author} {\bibinfo {author} {\bibfnamefont {L.}~\bibnamefont {Meunier}}, \bibinfo {author} {\bibfnamefont {H.}~\bibnamefont {Rakotoarison}}, \bibinfo {author} {\bibfnamefont {P.~K.}\ \bibnamefont {Wong}}, \bibinfo {author} {\bibfnamefont {B.}~\bibnamefont {Roziere}}, \bibinfo {author} {\bibfnamefont {J.}~\bibnamefont {Rapin}}, \bibinfo {author} {\bibfnamefont {O.}~\bibnamefont {Teytaud}}, \bibinfo {author} {\bibfnamefont {A.}~\bibnamefont {Moreau}},\ and\ \bibinfo {author} {\bibfnamefont {C.}~\bibnamefont {Doerr}},\ }\href@noop {} {\bibfield  {journal} {\bibinfo  {journal} {IEEE Transactions on Evolutionary Computation}\ }\textbf {\bibinfo {volume} {26}},\ \bibinfo {pages} {490} (\bibinfo {year} {2021})}\BibitemShut {NoStop}%
\bibitem [{\citenamefont {Raki{\'c}}\ \emph {et~al.}(1998)\citenamefont {Raki{\'c}}, \citenamefont {Djuri{\v{s}}i{\'c}}, \citenamefont {Elazar},\ and\ \citenamefont {Majewski}}]{rakic1998optical}%
  \BibitemOpen
  \bibfield  {author} {\bibinfo {author} {\bibfnamefont {A.~D.}\ \bibnamefont {Raki{\'c}}}, \bibinfo {author} {\bibfnamefont {A.~B.}\ \bibnamefont {Djuri{\v{s}}i{\'c}}}, \bibinfo {author} {\bibfnamefont {J.~M.}\ \bibnamefont {Elazar}},\ and\ \bibinfo {author} {\bibfnamefont {M.~L.}\ \bibnamefont {Majewski}},\ }\href@noop {} {\bibfield  {journal} {\bibinfo  {journal} {Applied optics}\ }\textbf {\bibinfo {volume} {37}},\ \bibinfo {pages} {5271} (\bibinfo {year} {1998})}\BibitemShut {NoStop}%
\bibitem [{\citenamefont {Frellsen}\ \emph {et~al.}(2016)\citenamefont {Frellsen}, \citenamefont {Ding}, \citenamefont {Sigmund},\ and\ \citenamefont {Frandsen}}]{frellsen_topology_2016}%
  \BibitemOpen
  \bibfield  {author} {\bibinfo {author} {\bibfnamefont {L.~F.}\ \bibnamefont {Frellsen}}, \bibinfo {author} {\bibfnamefont {Y.}~\bibnamefont {Ding}}, \bibinfo {author} {\bibfnamefont {O.}~\bibnamefont {Sigmund}},\ and\ \bibinfo {author} {\bibfnamefont {L.~H.}\ \bibnamefont {Frandsen}},\ }\href {https://doi.org/10.1364/OE.24.016866} {\bibfield  {journal} {\bibinfo  {journal} {Optics Express}\ }\textbf {\bibinfo {volume} {24}},\ \bibinfo {pages} {16866} (\bibinfo {year} {2016})},\ \bibinfo {note} {publisher: Optical Society of America}\BibitemShut {NoStop}%
\bibitem [{\citenamefont {Moreau}\ \emph {et~al.}(2012)\citenamefont {Moreau}, \citenamefont {Smaali}, \citenamefont {Centeno},\ and\ \citenamefont {Seassal}}]{moreau2012optically}%
  \BibitemOpen
  \bibfield  {author} {\bibinfo {author} {\bibfnamefont {A.}~\bibnamefont {Moreau}}, \bibinfo {author} {\bibfnamefont {R.}~\bibnamefont {Smaali}}, \bibinfo {author} {\bibfnamefont {E.}~\bibnamefont {Centeno}},\ and\ \bibinfo {author} {\bibfnamefont {C.}~\bibnamefont {Seassal}},\ }\href@noop {} {\bibfield  {journal} {\bibinfo  {journal} {Journal of Applied Physics}\ }\textbf {\bibinfo {volume} {111}} (\bibinfo {year} {2012})}\BibitemShut {NoStop}%
\bibitem [{\citenamefont {Christiansen}\ and\ \citenamefont {Sigmund}(2021)}]{christiansen2021compact}%
  \BibitemOpen
  \bibfield  {author} {\bibinfo {author} {\bibfnamefont {R.~E.}\ \bibnamefont {Christiansen}}\ and\ \bibinfo {author} {\bibfnamefont {O.}~\bibnamefont {Sigmund}},\ }\href@noop {} {\bibfield  {journal} {\bibinfo  {journal} {JOSA B}\ }\textbf {\bibinfo {volume} {38}},\ \bibinfo {pages} {510} (\bibinfo {year} {2021})}\BibitemShut {NoStop}%
\bibitem [{\citenamefont {Jiang}\ \emph {et~al.}(2020)\citenamefont {Jiang}, \citenamefont {Lupoiu}, \citenamefont {Wang}, \citenamefont {Sell}, \citenamefont {Hugonin}, \citenamefont {Lalanne},\ and\ \citenamefont {Fan}}]{jiang2020metanet}%
  \BibitemOpen
  \bibfield  {author} {\bibinfo {author} {\bibfnamefont {J.}~\bibnamefont {Jiang}}, \bibinfo {author} {\bibfnamefont {R.}~\bibnamefont {Lupoiu}}, \bibinfo {author} {\bibfnamefont {E.~W.}\ \bibnamefont {Wang}}, \bibinfo {author} {\bibfnamefont {D.}~\bibnamefont {Sell}}, \bibinfo {author} {\bibfnamefont {J.~P.}\ \bibnamefont {Hugonin}}, \bibinfo {author} {\bibfnamefont {P.}~\bibnamefont {Lalanne}},\ and\ \bibinfo {author} {\bibfnamefont {J.~A.}\ \bibnamefont {Fan}},\ }\href@noop {} {\bibfield  {journal} {\bibinfo  {journal} {Optics express}\ }\textbf {\bibinfo {volume} {28}},\ \bibinfo {pages} {13670} (\bibinfo {year} {2020})}\BibitemShut {NoStop}%
\end{thebibliography}%

\end{document}